\documentstyle[floats,tighten,preprint,eqsecnum,prd,aps,epsf]{revtex}

\begin{document}
\preprint{FERMILAB-PUB-97/412-T}
\draft

\hyphenation{author another created finan-cial paper re-commend-ed}

\title{The Self Energy of Massive Lattice Fermions}

\author{Bartholomeus P. G. Mertens}
\address{Enrico Fermi Institute and Department of Physics,
        University of Chicago, Chicago, Illinois}

\author{Andreas S. Kronfeld}
\address{Theoretical Physics Group,
		Fermi National Accelerator Laboratory, Batavia, Illinois}

\author{Aida X. El-Khadra}
\address{Physics Department, University of Illinois, Urbana, Illinois}

\date{December 20, 1997}

\maketitle 

\widetext       
\begin{abstract}
We address the perturbative renormalization of massive lattice 
fermions.
We derive expressions---valid to all orders in perturbation theory and 
for all values of the bare fermion mass---for the rest mass, the 
kinetic mass, and the wave-function renormalization factor.
We obtain the fermion's self energy at the one-loop level with a 
mass-dependent, $O(a)$ improved action.
Numerical results for two interesting special cases, the Wilson and 
Sheikholeslami-Wohlert actions, are given.
The mass dependence of these results smoothly connects the massless 
and infinite-mass limits, as expected.
Combined with Monte Carlo calculations our results can be employed to 
determine the quark masses in common renormalization schemes.
\end{abstract}


\narrowtext
\epsfverbosetrue


\section{Introduction}
For some time a goal of lattice QCD has been to determine the masses 
of the quarks.
To be precise, one would like to quote a value of $\bar{m}(\mu)$, the 
renormalized mass in the $\overline{\rm MS}$ scheme, at momentum
scale~$\mu$.
This is the convention most often used in the phenomenology of the 
Standard Model and in attempts to treat the Standard Model as the 
low-energy limit of a more fundamental theory.

In calculations of the hadron spectrum the bare (lattice) mass is a 
free parameter, which is adjusted to match experiment.
The $\overline{\rm MS}$ mass is related to the lattice mass via 
perturbation theory.
This relation is obtained by computing the quark's pole mass in 
dimensional regularization and in lattice perturbation theory, and 
then eliminating the pole mass.
For the light quarks the perturbative matching is well established at 
the one-loop level.
The results for the $\overline{\rm MS}$ mass (in the quenched 
approximation) form a consistent picture, at least when power-law 
lattice-spacing effects (from the underlying hadron masses) are taken 
into account~\cite{Gou97}.

For the charm and bottom quarks lattice artifacts may seem, at first 
glance, a greater worry, because~$m_qa$ (the quark mass in lattice 
units) is not necessarily small.
This is, however, not so.
Lattice artifacts take the form
\begin{equation}
	a\delta E = a^{s_n}b_n(m_qa)\langle\delta H_n\rangle
	\label{bH}
\end{equation}
where $\delta H_n$ is an operator whose matrix elements are 
(typically) insensitive or mildly sensitive to the heavy-quark mass,
and~$b(m_qa)$ is a $c$-number function.
In the static~\cite{Eic87,Eic90} and 
nonrelativistic~\cite{Cas86,Lep87} effective theories, which develop 
an expansion in~$1/m_q$, Eq.~(\ref{bH}) arises by design, and
the~$b_n$ are bounded for large~$m_qa$.
For Wilson-like actions Eq.~(\ref{bH}) also holds~\cite{KKM97}, and
the~$b_n$ are bounded for {\em all\/}~$m_qa$.
To obtain this result, it is essential to avoid expanding 
around~$m_qa=0$ or around~$m_q=\infty$ at every stage of the analysis.

The implication of Ref.~\cite{KKM97} for lattice perturbation theory 
is that the relationship between renormalized and bare quantities is 
needed for arbitrary~$m_qa$.
This paper examines mass and wave-function renormalization in the 
class of actions considered in Ref.~\cite{KKM97} and gives concrete 
results at the one-loop level for the Wilson~\cite{Wil77} and 
Sheikholeslami-Wohlert~\cite{She85} actions.
Results are available in the literature for 
$m_qa=0$~\cite{GHS84,Gab91,Cap97} and $m_q\to\infty$~\cite{Eic90}; as 
expected~\cite{KKM97}, the new results presented here smoothly connect 
the two limits.

The mass dependence of one-loop lattice perturbation theory has been 
considered before.
Results from nonrelativistic theories~\cite{Dav92,Mor93} provide us 
with cross checks, because (suitable combinations of) their results 
must agree with ours in the static limit,~$m_q\to\infty$.
At the other extreme, terms of order~$m_qa$, from expanding our 
general results around $m_q=0$, should recover the results of Sint and 
Weisz~\cite{Sin97}.

This paper is organized as follows:
Section~\ref{quark pole} discusses the pole in the lattice quark 
propagator for all~$m_qa$, to all orders in perturbation theory.
The action of Ref.~\cite{KKM97}, special cases of which are the Wilson 
and Sheikholeslami-Wohlert actions, is reviewed in Sec.~\ref{action}.
The main (all orders) results of Sec.~\ref{quark pole} are expanded to 
first order in~$g_0^2$ in Sec.~\ref{self-energy}, and expressions for 
one-loop Feynman diagrams are presented.
Some notation necessary for simplifying numerical evaluation of the 
one-loop diagrams is given in Sec.~\ref{numerical}.
Section~\ref {results} presents numerical results for the one-loop 
contributions to the rest mass, the kinetic mass, and the wave 
function renormalization factor.
(Some of the results have appeared 
previously~\cite{Mer94,Mer95,Mer97}.)
As usual, the dominant contributions come from tadpole diagrams;
the results of Sec.~\ref{results} are improved by ``mean-field 
theory''~\cite{Lep93} in Sec.~\ref{tadpoles removed}.
Some technical details are deferred to the Appendices.

The perturbation theory of this paper will be combined with Monte 
Carlo calculations of the quarkonium and heavy-light spectrum, to 
determine $\overline{\rm MS}$ masses $\bar{m}_{\text{ch}}$ and 
$\bar{m}_{\text{b}}$, in forthcoming publications~\cite{Kro97}.

\section{Renormalization to All Orders in Perturbation Theory}
\label{quark pole}
The objective of this section is to derive relations for mass and 
wave-function renormalization, for arbitrary values of~$m_qa$.
To do so, we shall assume only properties guaranteed in the lattice 
theory, and we shall not assume that the self energy is small.
By assuming less, we obtain more: our derivation succeeds not only 
for arbitrary mass, but to all orders in perturbation theory as well.
  
An outline of our analysis is as follows:
We start by anticipating the physical content of the quark propagator,
viewed as a function of three-momentum and time.
This provides a template from which one can read off the energy, as a 
function of three-momentum, and the wave-function renormalization 
factor.
We next write down a description of the free propagator and the 
self energy, as functions of the four-momentum, constrained only by 
symmetry and periodicity.
The description applies to all lattice theories with Wilson's 
discretization in time~\cite{Wil77}.
This includes the Wilson action~\cite{Wil77} (of course), the 
improvements of Ref.~\cite{She85} and of Ref.~\cite{KKM97}, and 
some nonrelativistic actions.
Then we Fourier transform the full propagator from four-momentum 
to time and three-momentum.
The result is a sum of terms, one for each pole in the 
momentum-space propagator.
Finally, we focus on the pole corresponding to the one-quark state, 
and read off the energy and wave-function renormalization factor from 
the template anticipated at the outset.

In our analysis it is unnecessary to assume that the three-momentum, 
the mass, or even the self energy itself, is small.
Indeed, the notion of a perturbation arises only to separate the 
inverse full propagator into a free part plus a self energy and to 
identify a one-quark state in the interacting theory.
Thus, our results for the pole are valid not only for all masses but 
also to all orders in the gauge interaction.
(They are not nonperturbative, because quark and gluon states have 
meaning, in nonabelian gauge theories, only within perturbation 
theory.)
Specializing to small three-momenta (in lattice units), we obtain the 
three main results of this section: all-orders formulae for the rest 
mass (Sec.~\ref{rm}), the kinetic mass (Sec.~\ref{km}), and the 
wave-function renormalization factor (Sec.~\ref{wfr}).
In subsequent sections, we specialize further, first to expressions 
for the one-loop self energy for the action given in 
Ref.~\cite{KKM97}, and later to numerical results for the Wilson and 
Sheikholeslami-Wohlert actions.

\subsection{The Quark Pole}
Because of confinement, the true states of QCD are hadrons, not quarks 
and gluons.
In perturbation theory, however, one may pretend that quark and gluon 
states exist.
Although the aim of this section is to relate the bare mass of lattice 
QCD to the perturbative pole mass, one should always view the pole 
mass as an intermediate step.
In a final application, the pole mass should be related to another 
regulator mass, such as the $\overline{\rm MS}$ mass, or to a 
genuine (hadronic) observable.

Mindful of the preceding caution, we proceed as if quarks and gluons 
are physical states.
The starting point is the quark two-point correlation function
\begin{equation}
	\langle \psi(t,\bbox{p}') \bar{\psi}(0,\bbox{p}) \rangle=
	(2\pi)^3\delta(\bbox{p}-\bbox{p}') G(t,\bbox{p}),
	\label{G definition}
\end{equation}
which defines~$G(t,\bbox{p})$.
The fields are those appearing in the functional integral: they are 
bare fields, in a fixed gauge chosen so that $G$ does not vanish 
trivially.
The field~$\bar{\psi}(0,\bbox{p})$ can create from the vacuum not just 
the one-quark state, but also states with extra gluons or 
extra $q\bar q$ pairs.
One thus anticipates
\begin{equation}
G(t,\bbox{p}) =
	{\cal Z}_2(\bbox{p}) e^{-E(\bbox{p})|t|} Q(\bbox{p})
		+ \int\frac{d^3k}{(2\pi)^3} 
	{\cal Z}_{qg}(\bbox{p},\bbox{k}) e^{-E_{qg}(\bbox{p},\bbox{k})|t|} 
		+ \cdots, \label{one quark G}
\end{equation}
where $E(\bbox{p})$ denotes the energy of the one-quark state with 
momentum~$\bbox{p}$, and~$E_{qg}$ the energy of states with a quark 
and a gluon.
(The quark-gluon states and the multi-particle states denoted by 
the ellipsis will not concern us further.)
If the $\gamma$ matrix $Q(\bbox{p})$ is normalized according to the 
condition given in Sec.~\ref{wfr}, then the 
residue~${\cal Z}_2(\bbox{p})$ is the square of the amplitude for the 
bare field to create a physical one-quark state.

With a Euclidean invariant cutoff the energy would satisfy 
$E(\bbox{p})=\sqrt{\bbox{p}^2+m^2}$, where~$m$ is the quark's 
``pole'' mass.
With a lattice cutoff, on the other hand, the mass shell is distorted.
To describe the distorted pole position in a systematic way, one can 
define a {\em rest} mass
\begin{equation}
	M_1=E(\bbox{0}),
	\label{M1 definition}
\end{equation}
a {\em kinetic} mass
\begin{equation}
	M_2=
	\left(\frac{\partial^2E}{\partial p_1^2}
	\right)^{-1}_{\bbox{p}=\bbox{0}},
	\label{M2 definition}
\end{equation}
and so on.
In general $M_1\neq M_2$, though as $a\to 0$ one should find 
$M_2\to M_1$.
Alternatively, at nonzero lattice spacing one can impose $M_2=M_1$ 
as a requirement on an improved action~\cite{KKM97}.

Similarly, with a Euclidean invariant cutoff the residue 
${\cal Z}_2(p)$ would be a function of $p^2$ only, evaluated on 
shell at $p^2=-m^2$.
It is thus a constant independent of~$\bbox{p}$; it is the 
wave-function renormalization factor~$Z_2$.
With a lattice cutoff, however, the $\bbox{p}$~dependence does not 
drop out of the residue, even on shell.
A reasonable definition of the wave-function renormalization factor 
is the residue at vanishing three-momentum
\begin{equation}
	Z_2 = {\cal Z}_2(\bbox{0}).
	\label{Z2 definition}
\end{equation}
Then $Z_2^{-1/2}\bar{\psi}$ creates the one-quark state with 
conventional (unit) normalization, at least for momenta much lower 
than the ultraviolet cutoff.

To obtain perturbative expressions for~$E(\bbox{p})$
and~${\cal Z}_2(\bbox{p})$, one starts in momentum space.
The inverse full propagator is written
\begin{equation}
G^{-1}(p)=G_0^{-1}(p)-\Sigma(p),
\end{equation}
where the self energy $\Sigma(p)$ is the sum of all one-particle 
irreducible graphs.
Given~$G_0$ and~$\Sigma$ one obtains the Fourier transform
\begin{equation}
	G(t,\bbox{p})=
	\int_{-\pi/a}^{\pi/a}\frac{dp_0}{2\pi}\,e^{ip_0t}G(p_0,\bbox{p})
	\label{G FT}
\end{equation}
and compares with Eq.~(\ref{one quark G}) to obtain $E(\bbox{p})$, 
${\cal Z}_2(\bbox{p})$, and~$Q(\bbox{p})$.

We now introduce suitably general expressions for the propagator.
For the lattice theories under consideration one can write the inverse 
free propagator as
\begin{equation}
	aG_0^{-1}(p)= i\kern+0.05em /\kern-0.7em K(p) + L(p),
	\label{free propagator}
\end{equation}
where
\begin{eqnarray}
	K_0(p_0,\bbox{p}) & = & \sin(p_0a),          \nonumber    \\
	\bbox{K}(p_0,\bbox{p}) & = & \bbox{K}(\bbox{p}), \label{K(p)} \\
	L(p_0,\bbox{p}) & = & \mu(\bbox{p})-\cos(p_0a).\nonumber
\end{eqnarray}
The exhibited $p_0$ dependence corresponds to Wilson's discretization, 
but the functions~$\bbox{K}(\bbox{p})$ and~$\mu(\bbox{p})$ depend on 
the action.%
\footnote{For nonrelativistic actions, one takes $\bbox{K}=\bbox{0}$ 
and projects out only the $\case{1}{2}(1+\gamma_0)$ component.}
We assume the action conserves parity, hence $\bbox{K}(\bbox{p})$ is 
an odd function of~$\bbox{p}$, and $\mu(\bbox{p})$ even.
The couplings are not explicit here, but reappear below as 
coefficients in the Taylor expansion of~$\bbox{K}$ and~$\mu$ around 
$\bbox{p}=\bbox{0}$.
For example, $\mu(\bbox{0})=1+m_0a$, where $m_0a$ is the bare mass.

We decompose the self energy into $\gamma$~matrices similarly to 
Eq.~(\ref{free propagator})
\begin{equation}
	a\Sigma(p) =
	i\sum_\rho \gamma_\rho A_\rho(p) \sin(p_\rho a) + C(p).
	\label{self energy}
\end{equation}
With a Euclidean invariant cutoff $A_\rho=A$ would be a single 
function for all~$\rho$, and $A$ and~$C$ would depend on $p^2$ only.
With a lattice cutoff, however, they are constrained only by 
(hyper)cubic symmetry.
For example, symmetry under parity implies that $A_\rho(p_0,\bbox{p})$ 
and $C(p_0,\bbox{p})$ are even functions of $p_0$ and~$\bbox{p}$; 
symmetry under cubic rotations implies that 
$A_1(0,p,0,0)=A_2(0,0,p,0)$; etc.
Furthermore, $A_\rho$ and~$C$ are periodic functions of $p_0$, with 
period~$2\pi/a$.
Below we make no assumptions about the self energy, except for these 
symmetry and periodicity properties.

Substituting Eqs.~(\ref{free propagator}) and~(\ref{self energy}) into 
Eq.~(\ref{G FT}), and adopting lattice units ($a=1$), one finds
\begin{equation}
	G(t,\bbox{p})=
	\int_{-\pi}^{\pi}\frac{dp_0}{2\pi}
	\frac{e^{ip_0t}N(p_0,\bbox{p})}%
	{2\nu(p_0,\bbox{p}) [\cosh{\cal E}(p_0,\bbox{p}) - \cos p_0]},
	\label{ugly FT}
\end{equation}
where $\nu(p_0,\bbox{p})=\mu(\bbox{p})-C(p_0,\bbox{p})$,
\begin{equation}
	N(p_0,\bbox{p}) = i\gamma_0\sin p_0[1-A_0(p_0,\bbox{p})]
	 + i\sum_i \gamma_i [K_i - \sin p_i A_i(p_0,\bbox{p})]
	 - [\nu(p_0,\bbox{p}) - \cos p_0], \label{prop num}
\end{equation}
and
\begin{eqnarray}
	2\nu\cosh{\cal E}(p_0,\bbox{p}) & = & 1 + \nu^2 + 
	\sum_i[K_i - A_i(p_0,\bbox{p}) \sin p_i]^2
	\nonumber  \\
	 & - & \sin^2p_0\{1-[1-A_0(p_0,\bbox{p})]^2\}.
	\label{pole position}
\end{eqnarray}
For $t\neq 0$ one can integrate over $p_0$ by changing variables to
$z=e^{ip_0\mathop{\rm sgn} t}$; then one has contour 
integration around the unit circle, and the integral is obtained 
through the residue theorem.

The integrand has a pole at $z=e^{-E}$, whenever $E$ solves the 
implicit equation
\begin{equation}
	\cosh E = \cosh{\cal E}(iE,\bbox{p}).
	\label{implicit energy}
\end{equation}
No more compact, general expression for $E$ exists but to set $p_0=iE$ 
in Eq.~(\ref{pole position}).
(At fixed order in perturbation theory one solves
Eqs.~(\ref{implicit energy}) and~(\ref{pole position}) iteratively.)
Solutions of Eq.~(\ref{implicit energy}) are parametrized by the 
three-momentum~$\bbox{p}$ and will be denoted~$E(\bbox{p})$.

In general, the integrand of Eq.~(\ref{ugly FT}) has several poles.  
In perturbation theory one assumes, however, that the self energy 
is a ``small'' correction.
Then the pole corresponding to the quark state must have a residue 
that does not vanish as $A_\rho,\;C\to 0$.
Poles corresponding to multi-particle states, on the other hand, must 
have residues that do vanish in the absence of an interaction.

Given an energy satisfying Eq.~(\ref{implicit energy}), one expands 
$\cosh{\cal E}(-i\ln z,\bbox{p})$ in $z$ around $e^{-E}$
\begin{eqnarray}
	\cosh{\cal E}(-i\ln z,\bbox{p}) &=& \cosh E 
		+ z^{-1} (z - e^{-E}) \dot{\cal E}(iE,\bbox{p}) \sinh E
		\label{E expansion} \\
		&+& O{\mathbf (}(z - e^{-E})^2{\mathbf )}, \nonumber
\end{eqnarray}
where 
\begin{equation}
	\dot{\cal E} = z \frac{d\,}{dz} {\cal E}(-i\ln z,\bbox{p}) =
	\frac{1}{i} \frac{d{\cal E}}{dp_0}(p_0,\bbox{p}).
	\label{dot E}
\end{equation}
In applying the residue theorem, the quadratic and higher-order terms 
drop out, and Eq.~(\ref{ugly FT}) becomes
\widetext\noindent
\narrowtext\noindent
\begin{eqnarray}
	G(t,\bbox{p}) &=&
	\frac{[1-A_0(iE,\bbox{p})]
	[\gamma_0\mathop{\rm sgn} t\sinh E
		-i\bbox{\gamma}\cdot\bbox{P}(\bbox{p})]
		+ \nu(iE,\bbox{p})-\cosh E}%
		{2\nu(iE,\bbox{p})[1+\dot{\cal E}(iE,\bbox{p})]
		\sinh E}	e^{-E|t|} 	\label{residues} \\
		&+& \text{other residues}, \nonumber
\end{eqnarray}
where, for brevity, $E=E(\bbox{p})$ and
\begin{equation}
	P_i(\bbox{p}) =
	\frac{K_i(\bbox{p}) - A_i(iE,\bbox{p})\sin p_i}%
	{1-A_0(iE,\bbox{p})}.
	\label{P}
\end{equation}
The chosen pole has a residue that remains when the interaction is 
turned off; thus, it corresponds to the one-quark state.
The ``other residues'' correspond to states other than the one-quark 
state and are disregarded from now on.

\subsection{Rest Mass $M_1$}
\label{rm}
To obtain an expression for the rest mass, one sets $p_0=iE(\bbox{p})$ 
in Eq.~(\ref{pole position}) and solves for $\nu$.
One finds
\begin{eqnarray}
	\nu(iE,\bbox{p}) 
	&\equiv& \mu(\bbox{p}) - C(iE,\bbox{p}) \nonumber \\
	&=&	\cosh E +
	(1-A_0) \sqrt{\sinh^2E - \bbox{P}^2}.
\label{coshE}
\end{eqnarray}
Setting $\bbox{p}=\bbox{0}$ in Eq.~(\ref{coshE}) yields the implicit 
equation
\begin{equation}\label{M1}
	e^{M_1} = 1 + m_0
			+ A_0(iM_1,\bbox{0})\sinh M_1 - C(iM_1,\bbox{0}),
\end{equation}
where the parameter
\begin{equation}
	m_0\equiv \mu(\bbox{0})-1	\label{bare mass}
\end{equation}
is the bare mass.
Equation~(\ref{M1}), expressing the rest mass to all orders in 
perturbation theory, is the first main result of this section.

For a massless fermion, the rest mass $M_1$ should vanish.
The critical bare mass, which induces $M_1=0$, is
\begin{equation}
	m_{0c}=C(0,\bbox{0};m_{0c}). \label{critical mass}
\end{equation}
The self energy depends on $m_0$ as a parameter, denoted here by the 
third argument of~$C$.
Since the lattice actions under consideration do not maintain explicit 
chiral symmetry, one expects $m_{0c}\neq 0$.
When applying Eq.~(\ref{M1}) it is useful to take care of this term 
once and for all and write
\begin{equation}\label{M1M0}
	e^{M_1} = 1 + M_0
			+ A_0(iM_1,\bbox{0})\sinh M_1 - \bar{C}(iM_1,\bbox{0})
\end{equation}
where $\bar{C}(iM_1,\bbox{0};m_0)=C(iM_1,\bbox{0};m_0)-m_{0c}$, and
\begin{equation}
	M_0 \equiv m_0-m_{0c}=\frac{1}{2\kappa}-\frac{1}{2\kappa_c}.
	\label{M0}
\end{equation}
In practice, one determines~$m_{0c}$ (or $\kappa_c$) nonperturbatively 
in Monte Carlo calculations and treats~$M_0$ (rather than~$m_0$) 
independently of~$g_0^2$ in perturbation theory.

\subsection{Kinetic Mass $M_2$}
\label{km}
From its definition [Eq.~(\ref{M2 definition})] the kinetic mass 
requires two derivatives with respect to~$p_1$.
Because the derivatives are applied to on-shell self-energy functions, 
the total derivative with respect to~$p_1$ includes an explicit part 
and an implicit part through the dependence on~$E$,
\begin{equation}
	\frac{d~}{dp_1}=\frac{\partial~}{\partial p_1} +
	i\frac{\partial E}{\partial p_1}
	 \frac{\partial~}{\partial p_0}.
\end{equation}
Differentiating Eq.~(\ref{coshE}) twice yields
\begin{equation}\label{M2}
\frac{e^{M_1}-A_0(iM_1,\bbox{0})\cosh M_1}{M_2}
	=	r_s\zeta + D(\bbox{0}) 
	+ \frac{[\zeta-A_1(iM_1,\bbox{0})]^2}%
			{[1-A_0(iM_1,\bbox{0})]\sinh M_1},
\end{equation}
where
\begin{equation}
	D(\bbox{p}) =
	\frac{d^2~}{dp_1^2}\left[A_0(iE(\bbox{p}),\bbox{p}) \sinh M_1
	-C(iE(\bbox{p}),\bbox{p}) \right].
	\label{D}
\end{equation}
The quantities
\begin{eqnarray}
	  \zeta  & \equiv &  K'_1(\bbox{0}), \label{zeta} \\
	r_s\zeta & \equiv & \mu''(\bbox{0}), \label{rs}
\end{eqnarray}
with primes denoting differentiation with respect to~$p_1$, are 
couplings parametrizing the action.
Equation~(\ref{M2}), expressing the kinetic mass to all orders in 
perturbation theory, is the second main result of this section.

Without the interaction that generates the self energy, one can solve 
Eq.~(\ref{M2}) in closed form: $M_2^{[0]}=m_2(M_1^{[0]})$, where 
\begin{equation}
	 m_2(m) = \frac{e^m\sinh m}{\zeta^2 + r_s\zeta \sinh m}.
	\label{tree M2}
\end{equation}
This expression suggests defining a kinetic-mass renormalization 
factor~$Z_{M_2}$ through
\begin{equation}
	Z_{M_2} = \frac{M_2}{m_2(M_1)}, \label{ZM2}
\end{equation}
which captures the radiative corrections to~$M_2$ not shared by~$M_1$.%
\footnote{Please note that it is the all-orders rest mass that appears
as the argument of the function~$m_2$.}
In perturbation theory a quark's kinetic and rest masses are on-shell 
observables.
By definition the ratio $Z_{M_2}$ is also an on-shell quantity.
It is, therefore, a useful diagnostic of cutoff effects in the 
continuum limit.
Indeed, 
\begin{equation}
	\lim_{M_1a\to 0}Z_{M_2}=1
\end{equation}
to {\em all\/} orders in perturbation theory.

For $M_1a\neq 0$ the rest and kinetic masses are not, as a rule, equal.
To construct a mass-dependent improved action, as in 
Ref.~\cite{KKM97}, one sets $M_2=M_1$ in Eq.~(\ref{M2}) and solves for 
a condition on the coupling~$\zeta$, parametrized by~$r_s$.
In the following we do not, however, assume that such conditions have 
been imposed.

\subsection{Wave-function Renormalization $Z_2$}
\label{wfr}
Comparing Eq.~(\ref{residues}) with Eq.~(\ref{one quark G}) one 
identifies the $\gamma$ matrix function of $\bbox{p}$ that multiplies 
$e^{-E(\bbox{p})|t|}$ as ${\cal Z}_2Q$.
To fix the normalization of~$Q$, first write 
$Q=\mathop{\rm sgn} t\gamma_0 Q_0-{i\bbox{\gamma}\cdot\bbox{Q}}+R$.
(Note that $Q_0^2-\bbox{Q}^2-R^2=0$ on shell.)
In the canonical normalization $Q_0=\case{1}{2}$.
Thus,
\begin{equation}
	{\cal Z}_2(\bbox{p}) = \frac{1-A_0(iE,\bbox{p})}%
		{\nu(iE,\bbox{p}) 
		[1+\dot{\cal E}(i E,\bbox{p}) ]}.
	\label{Z2(A0,dotE)}
\end{equation}
A more explicit expression may be obtained by using
Eqs.~(\ref{pole position}) and~(\ref{implicit energy}) to eliminate 
$\nu(iE,\bbox{p})$ and 
$\nu\dot{\cal E}(iE,\bbox{p})$.
One finds
\begin{eqnarray}
	{\cal Z}_2(\bbox{p})^{-1} 
	&=&	(1-A_0)\cosh E + \sqrt{\sinh^2E-\bbox{P}^2} 
	+	\dot{A}_0\sinh E \label{Z2(A,C)} \\
	&-&	\dot{C}\sqrt{1-\bbox{P}^2/\sinh^2E}  
	-	\sum_j \frac{\dot{A}_j P_j \sin p_j}{\sinh E}, \nonumber
\end{eqnarray}
with all self-energy functions evaluated at $(iE(\bbox{p}),\bbox{p})$.
Setting $\bbox{p}=\bbox{0}$ to obtain the wave-function 
renormalization factor, as discussed above, one finds
\begin{eqnarray}
	Z_2^{-1} &=& {\cal Z}_2(\bbox{0})^{-1} \nonumber \\
		&=&	e^{M_1} - A_0\cosh M_1 + \dot{A}_0 \sinh M_1 - \dot{C},
		\label{Z2}
\end{eqnarray}
where all self-energy functions are evaluated at $(iM_1,\bbox{0})$.
Note that every term on the right-hand side of Eq.~(\ref{Z2}) is of 
order~$e^{M_1}$ in the large mass limit, just as at tree level.
Equation~(\ref{Z2}), expressing the wave-function renormalization 
factor for arbitrary values of~$M_1a$, to all orders in perturbation 
theory, is the third, and final, main result of this section.

In most gauges, the wave-function renormalization factor is infrared 
divergent.
The divergence cancels against vertex renormalization factors, when a 
physical combination, such as the full renormalization of a current, 
is considered.

\section{The Lattice Action}
\label{action}
We consider the action $S=S_0+S_B+S_E$ of Ref.~\cite{KKM97}, namely
\begin{eqnarray}
  S_0	&=& m_0 a^4\sum_x \bar{\psi}(x)\psi(x) 
		+ \case{1}{2}a^4\sum_x 
			\bar{\psi}(x)[(1+\gamma_0)D_0^- - (1-\gamma_0)D_0^+]\psi(x)
			\nonumber \\
		&+& \zeta a^4\sum_x
			\bar{\psi}(x)\bbox{\gamma}\cdot\bbox{D}\psi(x) 
		-  \case{1}{2} r_s\zeta a^5\sum_x
			\bar{\psi}(x)\triangle^{(3)}\psi(x), \label{S0}
\end{eqnarray}
where the covariant difference operators are
\begin{equation}
	aD_0^{\pm}\psi(x)= \pm[U_{\pm 0}(x)\psi(x\pm a\hat{0})-\psi(x)],
	\label{D_0}
\end{equation}
\begin{equation}
	aD_i\psi(x) = \case{1}{2}[U_i\psi(x+a\hat{\imath})
		- U_{-i}\psi(x-a\hat{\imath})],	\label{D_i}
\end{equation}
and 
\begin{equation}
	a^2\triangle^{(3)}\psi(x)=\sum_i
	\case{1}{2}\big[U_i\psi(x+a\hat{\imath})
	+U_{-i}\psi(x-a\hat{\imath})-2\psi(x)\big] .
	\label{lat Lap}
\end{equation}
The action $S_0$ has cutoff artifacts of order~$a$, which can be 
canceled by the interactions
\begin{equation}\label{SB}
S_B=-\case{i}{2} a^5c_B\zeta\sum_x
\bar{\psi}(x)\bbox{\Sigma}\cdot\bbox{B}(x)\psi(x),
\end{equation}
\begin{equation}\label{SE}
S_E=-\case{1}{2} a^5c_E\zeta\sum_x
\bar{\psi}(x)\bbox{\alpha}\cdot\bbox{E}(x)\psi(x),
\end{equation}
for appropriate adjustments of~$c_B$ and~$c_E$.
The chromomagnetic and chromoelectric fields are given in 
Ref.~\cite{KKM97}.

Special cases of this action are the Wilson action~\cite{Wil77}, which 
sets $r_s=\zeta=1$, $c_B=c_E=0$; and the Sheikholeslami-Wohlert 
action~\cite{She85}, which sets $r_s=\zeta=1$,
$c_B=c_E\equiv c_{\text{SW}}$.
But to remove lattice artifacts for arbitrary masses,
the couplings~$r_s$, $\zeta$, $c_B$ and~$c_E$
must be taken to depend on $m_0a$~\cite{KKM97}.
For the purposes of this paper, however, the additional couplings are 
taken as free parameters.

We note here the elements of the free propagator introduced in 
Eq.~(\ref{K(p)}).
  From Eq.~(\ref{S0}) one finds 
\begin{eqnarray}
K_i(\bbox{p})&=&\zeta\sin p_ia, \label{KKM K} \\ 
\mu(\bbox{p})&=&1+m_0a+\case{1}{2}r_s\zeta\hat{\bbox{p}}^2a^2,
\label{KKM L}
\end{eqnarray}
where $\hat{p}_i=(2/a)\sin\case{1}{2}p_ia$.
Thus, the notation for the couplings coincides with that in 
Eqs.~(\ref{bare mass}), (\ref{zeta}), and~(\ref{rs}).

As the mass tends to infinity, all actions described by $S_0+S_B+S_E$
lead, up to an unphysical factor, to the same quark propagator---a 
Wilson line.
Perturbative corrections to masses and vertices must respect this
universal static limit, and, therefore, they must tend to a universal 
value.
This limiting behavior is a helpful check.

\section{The Self Energy to One Loop}
\label{self-energy}
The analysis of Sec.~\ref{quark pole} is valid to all orders in 
perturbation theory.
We now develop expansions in~$g_0^2$ for the main results, 
concentrating on the one-loop approximation.
We also present our expressions for the one-loop self energy.

\subsection{Perturbative Series}
\label{series}

In perturbation theory the self energy is expanded
\begin{equation}
	\Sigma(p) = \sum_{l=1}^\infty g_0^{2l}\Sigma^{[l]}(p)
	\label{Sigma series},
\end{equation}
and similarly for the functions~$A_\rho$ and~$C$.
As a consequence, the rest mass has an expansion
\begin{equation}
	M_1 = \sum_{l=0}^\infty g_0^{2l} M_1^{[l]}.
	\label{M1 series}
\end{equation}
where the tree level $M_1^{[0]}=\log(1+M_0)$.
The one-loop coefficient, from Eq.~(\ref{M1}), is
\begin{equation}
	M_1^{[1]}=\left[A_0^{[1]}(iM_1^{[0]},\bbox{0})\sinh M_1^{[0]}
			 -\bar{C}^{[1]}(iM_1^{[0]},\bbox{0})\right] e^{-M_1^{[0]}}.
	\label{M1[1]}
\end{equation}
In the massless limit 
\begin{equation}
	\frac{M_1^{[1]}}{M_1^{[0]}}=A_0^{[1]}-
	\left.\frac{\partial C^{[1]}}{\partial m_0}\right|_{m_0=0},
	\label{M1 continuum}
\end{equation}
which is the same result as in massless derivations~\cite{GHS84}.
In the static limit, $A_0$ tends to a constant and $C$ to 
$e^{M_1}$ times a constant.
Thus, $M_1^{[1]}(\infty)$ is finite.
Moreover, it is the same for all actions under consideration.

The kinetic-mass renormalization factor has an expansion
\begin{equation}
	Z_{M_2} = 1 + \sum_{l=1}^\infty g_0^{2l} Z_{M_2}^{[l]}
	\label{ZM2 series}
\end{equation}
The one-loop coefficient, from Eq.~(\ref{M2}), is
\widetext
\begin{equation}
Z_{M_2}^{[1]} 
	= \frac{2\zeta A_1^{[1]}(iM_1^{[0]},\bbox{0})
	-	\zeta^2 A_0^{[1]}(iM_1^{[0]},\bbox{0})
	-	D^{[1]}(\bbox{0})\sinh M_1^{[0]}}%
	   {\zeta^2+r_s\zeta\sinh M_1^{[0]}} 
	- A_0^{[1]}(iM_1^{[0]},\bbox{0}) \cosh M_1^{[0]} e^{-M_1^{[0]}}.
	\label{ZM2[1]}
\end{equation}
\narrowtext\noindent
In the massless limit, $Z_{M_2}^{[1]}$ vanishes at least as fast as 
$(Ma)^2\ln(Ma)$.
In the static limit
$Z_{M_2}^{[1]}\to-D^{[1]}/(r_s\zeta)-\case{1}{2}A_0^{[1]}$,
again finite.
Moreover, it too is the same for all actions under consideration, and 
it can be compared to the same combination in nonrelativistic QCD.

At tree level many have noticed that the quark propagator's residue
${\cal Z}_2(\bbox{0})=e^{-M_1}$.
This dominant (large) mass dependence persists in individual loop 
diagrams~\cite{KKM97} and, as shown in Eq.~(\ref{Z2}), to all orders.
To isolate the subleading mass dependence of the wave-function 
renormalization, we develop the expansion as follows:
\begin{equation}
	e^{M_1}Z_2 = 1 + \sum_{l=1}^\infty g_0^{2l} Z_2^{[l]},
	\label{Z2 series}
\end{equation}
with the physical (all-orders) rest mass in the exponent on the 
left-hand side.
We apologize for a notation in which $Z_2^{[l]}$ is {\em not\/} the 
perturbative coefficient of $Z_2$, but this way the~$Z_2^{[l]}$ have 
only mild mass dependence.
The one-loop coefficient, from Eq.~(\ref{Z2}), is
\begin{equation}
	Z_2^{[1]} = \left[A_0^{[1]}(iM_1^{[0]},\bbox{0})\cosh M_1^{[0]}
			  - \dot{A}_0^{[1]}(iM_1^{[0]},\bbox{0})\sinh M_1^{[0]}
			  + \dot{C}^{[1]}(iM_1^{[0]},\bbox{0})\right]
              e^{-M_1^{[0]}}.
	\label{Z2[1]}
\end{equation}
In the massless limit, the right-hand side reduces to the well-known 
result~$A_0^{[1]}-M\dot{A}_0^{[1]}+\dot{C}^{[1]}$.
(As $M\to 0$, $M\dot{A}^{[1]}\to\text{const}$.)
In the static limit, each term on the right-hand side of approaches
a (universal) value, and thus the sum~$Z_2^{[1]}$ does too.

\subsection{One-Loop Diagrams}
\label{diagrams}
\begin{figure}[btp]
	\begin{center}
		\epsfxsize=0.75\textwidth
		\epsfbox{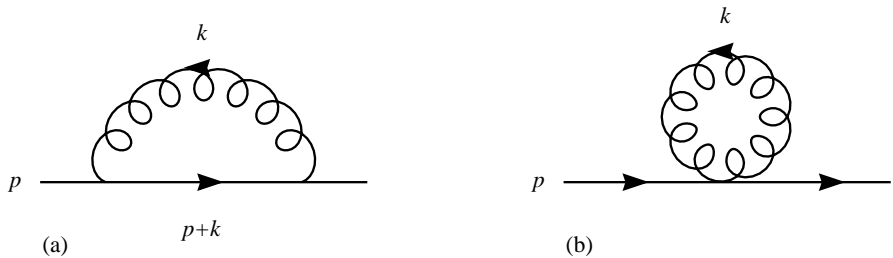}
	\end{center}
	\caption{Feynman diagrams for the one-loop self-energy:
	(a)~rainbow, (b)~tadpole.}
	\label{fig:self-energy}
\end{figure}

It is straightforward to derive Feynman rules for $S_0+S_B+S_E$.
They are listed in Appendix~\ref{feynman rules}.
To one loop the self energy is given by the Feynman diagrams in 
Fig.~\ref{fig:self-energy}.
We use the Wilson gauge action~\cite{Wil74}.
The rainbow diagram, Fig.~\ref{fig:self-energy}a, represents the 
contribution
\begin{equation}
	\Sigma^{[1]}_{\text{(a)}}(p) =
		C_F\int\frac{d^4k}{(2\pi)^4}
		\frac{F(p,k)}{\hat{k}^2[K^2(p+k) + L^2(p+k)]},
	\label{Sigma 1}
\end{equation}
with $K_\rho$ and $L$ specified by Eqs.~(\ref{K(p)}), (\ref{KKM K}), 
and~(\ref{KKM L}).
The numerator~$F(p,k)$, which follows from $\gamma$-matrix algebra, is 
given in Appendix~\ref{numerators}.
The color factor $C_F=(N_c^2-1)/(2N_c)$ [$=4/3$ for SU(3)].
The tadpole diagram, Fig.~\ref{fig:self-energy}b, represents a much 
simpler contribution; in Feynman gauge
\begin{equation}
	\Sigma^{[1]}_{\text{(b)}}(p) =
		\case{1}{2}C_F
		[i\gamma_0 \sin p_0 + i\zeta\bbox{\gamma}\cdot\sin\bbox{p} 
		- \cos p_0 - r_s \zeta\sum_i\cos p_i] 
		\int\frac{d^4k}{(2\pi)^4}\frac{1}{\hat{k}^2},
	\label{Sigma 1 tadpole}
\end{equation}
which is independent of $c_B$ and $c_E$.

On shell the total one-loop self energy is gauge independent.
Therefore, the one-loop radiative corrections~$M_1^{[1]}$ 
and~$Z_{M_2}^{[1]}$ are gauge independent, as one expects.
Derivatives of the (off-shell) self energy do, however, depend on the 
gauge parameter.
Therefore, the wave-function renormalization factor does depend on the 
gauge; below we present the result in Feynman gauge.
Note, furthermore, that terms arising from~$S_B$ and~$S_E$ are gauge 
independent (at the one-loop level).

By infrared power counting, one expects the one-loop self 
energy~$\Sigma^{[1]}$ to contain logarithmic nonanalyticity 
as~$Ma\to 0$.
The leading nonanalyticity is the same as in a Pauli-Villars regulator.
The latter amounts to using the gluon propagator
\begin{equation}\label{PV prop}
\Delta(k^2)=\frac{1}{k^2+\lambda^2}-\frac{1}{k^2+a^{-2}},
\end{equation}
where the gluon mass~$\lambda$ serves as an infrared regulator.
As with the lattice calculation, we do not necessarily assume 
that~$Ma$ is small, and we usually set $a=1$.
Thus,
\begin{equation}
 	\Sigma^{[1]}_{\text{PV}}(\kern+0.1em /\kern-0.55em p;m)=
 	-C_F\int \frac{d^4k}{16\pi^2}\,
 	\frac{2i(\kern+0.1em /\kern-0.55em p+
 			 \kern+0.1em /\kern-0.55em k)+4m}{(p+k)^2+m^2} \Delta(k^2).
	\label{PV Sigma}
\end{equation}
As in Eq.~(\ref{self energy}) we write
\begin{equation}
	\Sigma^{[1]}_{\text{PV}}=i\kern+0.1em /\kern-0.55em p
		A^{[1]}_{\text{PV}}	+ C^{[1]}_{\text{PV}}.
\end{equation}
Below it is sometimes convenient to set
\begin{equation}
	C^{[1]}_{\text{PV}}(p^2;m)=mB^{[1]}_{\text{PV}}(p^2;m),
	\label{C=mB}
\end{equation}
because the Pauli-Villars regulator does not break chiral symmetry.
Equation~(\ref{PV prop}) specifies a Euclidean-invariant cutoff,
so~$A^{[1]}_{\text{PV}}$ and~$B^{[1]}_{\text{PV}}$ are functions 
of~$p^2$, rather than of~$(p_0,\bbox{p})$.
Logarithms arise in Eq.~(\ref{PV Sigma}) from the region $k^2\ll 1$.

Below we exploit the similarities of the two regulators to isolate 
analytically terms of the form $\ln Ma$ and $Ma\ln Ma$.

\section{Summary of Numerical Methods}
\label{numerical}
In this section we outline the numerical procedures used to evaluate 
the loop integrals.
In particular, we isolate from~$M_1^{[1]}$, $Z_{M_2}^{[1]}$, 
and~$Z_2^{[1]}$ parts that are easy to compute numerically.
The notation introduced is needed below to obtain the one-loop 
coefficients, as a function of mass, from the tables in 
Sec.~\ref{results}.
Some other technical details are deferred to appendices.

The remainder of this paper focuses on one-loop renormalization for 
the action used in Monte Carlo calculations, namely $r_s=\zeta=1$ 
and $c_B=c_E\equiv c_{\text{SW}}$.
For future flexibility it is useful to classify the results as a
(second-order) polynomial in~$c_{\text{SW}}$, for example
\begin{equation}
	A_0^{[1]}=A_0^{[1](0)}+
	c_{\text{SW}}A_0^{[1](1)}+c_{\text{SW}}^2A_0^{[1](2)}.
	\label{c poly}
\end{equation}

\subsection{Numerical Integration}
Because one must analytically continue the self energy from 
real~$p_0$ to imaginary~$iE(\bbox{p})$, it proved wise to carry out 
the integration over~$k_0$ analytically.
A full discussion of this technicality is in 
Appendix~\ref{integration}.
Here we focus on the remaining integration over~$\bbox{k}$.

In numerical evaluation of the lattice integrals, it is helpful to 
compute the difference between the lattice and Pauli-Villars 
regulated integrals.
The key is to subtract the two integrands in momentum space and to add
the analytical expression for the Pauli-Villars integral afterwards.
Then the numerical integration package does not need to uncover the 
logarithmic singularities.
At small~$\bbox{k}$ the lattice integrands take the form
\begin{equation}
	\frac{p(M_1^{[0]})}{2\sinh M_1^{[0]}|\bbox{k}|^2}
\end{equation}
The denominator of the Pauli-Villars integrand takes the same form if 
one sets $m=\sinh M_1^{[0]}$ in Eq.~(\ref{PV Sigma}).
If one also multiplies the Pauli-Villars integrand by the 
function~$p$, the numerical integration package has an even easier 
job, because then the subtraction removes contributions of the 
form~$Ma\ln Ma$ (for small~$Ma$).

Thus, let
\begin{eqnarray}
	a_0^{[1](n)} &=&
	A_0^{[1](n)}-p_{A_0}^{(n)}(M_1^{{[0]}})A^{[1]}_{\text{PV}},
	\label{A0 sub} \\
	a_1^{[1](n)} &=&
	A_1^{[1](n)}-p_{A_1}^{(n)}(M_1^{{[0]}})A^{[1]}_{\text{PV}},
	\label{A1 sub} \\
	c^{[1](n)} &=&
	\bar{C}^{[1](n)}-p_{C}^{(n)}(M_1^{{[0]}})C^{[1]}_{\text{PV}}.
	\label{C sub}
\end{eqnarray}
The Pauli-Villars subtractions on the right-hand sides are done on the 
integrands.
They are needed for the Wilson-action ($n=0$) contribution 
only.
We find 
\begin{eqnarray}
	p_{A_0}^{(0)}(M) & = & \case{1}{2}(3e^{-M}-e^M),
	\label{pA0}  \\
	p_{A_1}^{(0)}(M) & = & e^{-M}(1-\sinh M),
	\label{pA1}  \\
	p_{C}^{(0)}(M)   & = & \case{1}{4}(3e^{-M}+e^M),
	\label{pC}
\end{eqnarray}
and $p_{\Sigma}^{(n)}=0$ otherwise.
The subtraction $C(iM_1;m_0)-C(0;m_{0c})=\bar{C}$ is also done on the
integrand.

The kinetic mass requires also the function~$D$, defined in 
Eq.~(\ref{D}).
The total derivative with respect to~$p_1$ acts on the 
energy~$E(\bbox{p})$ and the explicit $p_1$-dependence.
Each generates a severely divergent peak in the integrand 
of~$D^{[1](0)}$ at somewhat different infrared locations.
After integrating, the infrared divergences cancel exactly.
To make numerical integration easier, it is again prudent to let
\begin{equation}
	d^{[1](0)} =
	D^{[1](0)}-p_D(M_1^{[0]})D^{[1]}_{\text{PV}} \label{D sub},
\end{equation}
to cancel the peaks against integrands from the Pauli-Villars 
regulator.
The function~$p_D$ is not needed below, however, because the
integral~$D^{[1]}_{\text{PV}}$ vanishes identically, but we take
$p_D(M)=e^M/[\cosh M+\case{1}{2}(e^{-M}-1)\sinh M]^3$.

For the wave-function renormalization factor we compute
\begin{equation}
	z_2^{[1](n)} 
		= \left[a_0^{[1](n)}\cosh M_1^{[0]} \right. 
		- \left. \dot{a}_0^{[1](n)}\sinh M_1^{[0]} 
		 + \dot{c}^{[1](n)}\right] e^{-M_1^{[0]}}, \label{Z2 sub}
\end{equation}
where $\dot{a}_0^{[1](n)}$ and $\dot{c}^{[1](n)}$ are defined 
analogously to~$a_0^{[1](n)}$ and~$c^{[1](n)}$.
The $z_2^{[1](n)}$ are finite as $\lambda\to 0$ and as $Ma\to 0$.
In any physical quantity the wave-function renormalization is combined 
with vertex corrections in an infrared-finite way. 
Thus, $z_2^{[1]}$ is a suitable synopsis of the lattice 
renormalization.

The subtractions permit a numerical evaluation of $a_0^{[1](n)}$, 
$a_1^{[1](n)}$, $c^{[1](n)}$, $d^{[1](n)}$, and $z_2^{[1](n)}$ with 
gluon mass $\lambda=0$.
With the subtracted integrals in hand, the lattice self energy can be 
reconstructed with the closed forms for
$A^{[1]}_{\text{PV}}$, $B^{[1]}_{\text{PV}}$, 
$\dot{A}^{[1]}_{\text{PV}}$, and $\dot{B}^{[1]}_{\text{PV}}$, 
given in Appendix~\ref{ABPV}.

\subsection{Chebyshev Approximation}
With the adaptive integration routine {\sc vegas} we evaluate 
$a_0^{[1](n)}$, $a_1^{[1](n)}$, $c^{[1](n)}$, $d^{[1](n)}$, 
and~$z_2^{[1](n)}$ at 51 values of the mass, chosen such that
\begin{equation}
	\tanh M_1^{[0]}=\case{1}{2}(1+x_k),\quad k=0,~50,
	\label{Chebyshev masses}
\end{equation}
where the $x_k=\cos[\pi(k+\case{1}{2})/51]$ are the zeroes of the
51st Chebyshev polynomial.
This procedure allows us to combine the individual evaluations into a 
Chebyshev approximation to the exact result.
Let
\begin{equation}
	f_j=\frac{2}{N}\sum_{k=0}^{N-1} f(x_k) T_j(x_k),
	\label{Chebyshev coefficients}
\end{equation}
where $N=51$ and $T_j(x)=\cos(j\cos^{-1}x)$ is the $j$th Chebyshev 
polynomial.
Then
\begin{equation}
	f(x)\approx \case{1}{2}f_0 + \sum_{k=1}^{m-1} f_k T_k(x)
	\label{Chebyshev approx}
\end{equation}
is (expected to be) a good approximation even for $m<N$.
Since $|T_j(x)|\le 1$ the utility of the approximation can be 
ascertained from inspecting the~$f_j$.
Section~\ref{results} gives tables with the first several coefficients 
of Chebyshev expansions.
All 51 values of~$f(x_k)$ and~$f_j$ are available on the WorldWideWeb 
at
{\tt http://www-theory.fnal.gov/people/ask/self-energy/}.

The lattice self-energy functions have also been obtained over a wide 
range of masses~\cite{Mer97} without the Pauli-Villars subtractions.
The results agree, of course, but with the subtractions one can reach 
better precision more quickly.

\section{One-Loop Results}
\label{results}
\subsection{Critical Bare Mass $m_{0c}$}
\label{one-loop m0}
For completeness we give here our result for the one-loop bare mass 
that makes the physical masses vanish:
\begin{equation}
	m_{0c}^{[1]} = C^{[1]}(0,\bbox{0}) =
	C_F\left[-0.325714(5)
		 +    0.086964(9) c_{\text{SW}}^{ } 
		 +    0.036190(2) c_{\text{SW}}^2 \right].
	\label{m0 one-loop result}
\end{equation}
Errors on the least significant digit(s) from numerical integration 
are given in parentheses.
Equation~(\ref{m0 one-loop result}) agrees with published values:
$-C_F0.325789(3)$ at $c_{\text{SW}}=0$~\cite{GHS84} and
$-C_F0.20(1)$ at $c_{\text{SW}}=1$~\cite{Gab91}.
The individual coefficients of~$c_{\text{SW}}$ agree with 
Ref.~\cite{Cap97}.

\subsection{Rest Mass $M_1$}
\label{one-loop M1}
\begin{figure}[btp]
\begin{center}
	\epsfxsize=0.45454545\textwidth
	\epsfbox{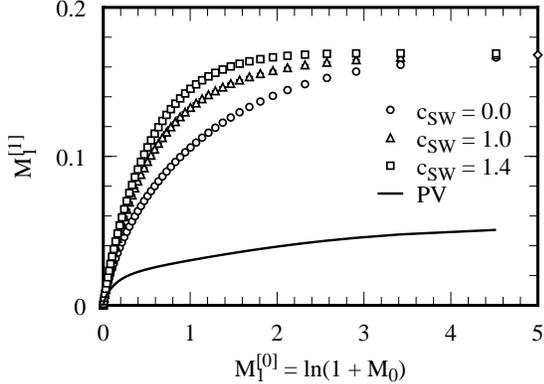}
\end{center}
\caption[fig:M1-1]{Plot of $M_1^{[1]}$ vs.\ $M_1^{[0]}$ for 
$c_{\text{SW}}=0$~(circles), 1~(triangles), and 1.4~(squares). 
The diamond marks the static limit~\cite{Eic90}.
To illustrate that the overall shape is not an artifact of lattice
field theory, the curve shows~$M_{\text{PV}}^{[1]}$.
Here $C_F=4/3$.}\label{fig:M1-1}
\end{figure}

Figure~\ref{fig:M1-1} shows the one-loop correction to the rest 
mass~$M_1$.
We present results for three values of~$c_{\text{SW}}$:
0 (Wilson action), 1 (tree-level improvement), 
and 1.4 (a typical mean-field estimate of $c_{\text{SW}}$).
As expected, $M_1^{[1]}$ smoothly connects to the massless and static 
limits.
As $M_1\to\infty$ all curves approach the same limiting value, 
$M_1^{[1]}(\infty)=C_F0.1261(2)$, which agrees with the value 
$C_F0.1263(1)$ obtained directly in the static limit~\cite{Eic90}.
We are able to reproduce the result of Ref.~\cite{Kur97}, which
considers only the Wilson action, if we omit the tadpole diagram's
contribution~$\Sigma^{[1]}_{\text{(b)}}$.

\begin{figure}[btp]
\begin{center}
	\epsfxsize=0.45454545\textwidth
	\epsfbox{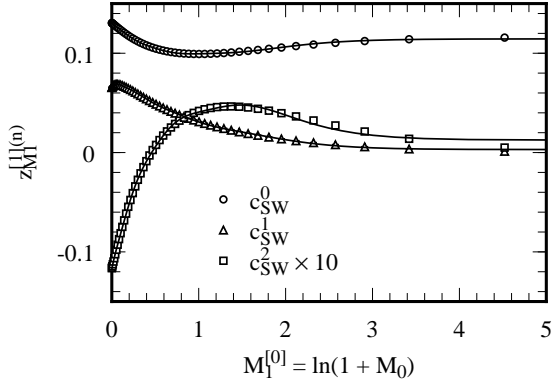}
\end{center}
\caption[fig:ZM1-1]{Plot of $z_{M_1}^{[1]}$ vs.\ $M_1^{[0]}$ as 
coefficients of~$c_{\text{SW}}$.
The curves indicate the fifteen-term Chebyshev interpolation. 
Note that~$z_{M_1}^{[1](2)}$ (the coefficient of~$c_{\text{SW}}^2$) is 
multiplied by ten.  Here $C_F=4/3$.}\label{fig:ZM1-1}
\end{figure}

For $M_1a<1$ the additive form is not illuminating, because there
$M_1^{[1]}\propto M_1^{[0]}$.
It is convenient to define the rest-mass renormalization factor%
\footnote{The denominator $\tanh M_1^{[0]}$ is handy because 
$\tanh m=m+O(m^3)$ at small~$m$, yet $\tanh\infty=1$.}
\begin{equation}
	Z_{M_1} \equiv \frac{M_1a}{\tanh M_1^{[0]}a}. \label{ZM1}
\end{equation}
To build the one-loop renormalization factor from the {\sc vegas} 
integrals let
\begin{equation}
	z_{M_1}^{[1](n)} = e^{-M_1^{[0]}}\left(
	a_0^{[1](n)}\cosh M_1^{[0]} - c^{[1](n)}\coth M_1^{[0]}\right),
	\label{zM1}
\end{equation}
and then
\begin{equation}
	Z_{M_1}^{[1](n)} = z_{M_1}^{[1](n)} 
		+ e^{-M_1^{[0]}}\cosh M_1^{[0]}
		 \left[p_{A_0}^{(n)}(M_1^{{[0]}}) A^{[1]}_{\text{PV}} -
	    p_{C}^{(n)}(M_1^{{[0]}})   B^{[1]}_{\text{PV}}
	\right].  \label{ZM1[1]}
\end{equation}
Figure~\ref{fig:ZM1-1} shows the~$z_{M_1}^{[1](n)}$ and
Table~\ref{tbl:zM1-Cheb} contains the first 15 coefficients of their
Chebyshev expansions.
\begin{table}[tbp]
\centering
\caption[tbl:zM1-Cheb]{Coefficients of Chebyshev polynomials~$T_j(x)$ 
for~$z_{M_1}^{[1](n)}$.  Here $C_F=4/3$.}\label{tbl:zM1-Cheb}
\begin{tabular}{c*{3}{r@{.}l}}
$j$	& \multicolumn{2}{c}{$z_{M_1}^{[1](0)}$}
	& \multicolumn{2}{c}{$z_{M_1}^{[1](1)}$}
	& \multicolumn{2}{c}{$z_{M_1}^{[1](2)}$} \\ 
\hline
 0  &    0&222802     &   0&0868169    & $-0$&00278643  \\
 1  & $-0$&0117152    & $-0$&0300771   &   0&00744513   \\
 2  &    0&00883238   & $-0$&0052349   & $-0$&00297984  \\
 3  &    0&00216047   & $-0$&000493063 & $-0$&00039969  \\
 4  &    0&00141579   & $-0$&00245913  & $-0$&000478208 \\
 5  &    0&000994781  & $-0$&000371327 & $-0$&000358493 \\
 6  &    0&000590983  & $-0$&000965165 & $-0$&000237546 \\
 7  &    0&000467076  & $-0$&000311904 & $-0$&000198283 \\
 8  &    0&000321860  & $-0$&000497263 & $-0$&000147608 \\
 9  &    0&000256113  & $-0$&000225545 & $-0$&000123028 \\
10  &    0&000195137  & $-0$&000302366 & $-9$&86053$\times10^{-5}$ \\
11  &    0&000159702  & $-0$&000161754 & $-8$&34899$\times10^{-5}$ \\
12  &    0&000125988  & $-0$&000203734 & $-6$&96978$\times10^{-5}$ \\
13  &    0&000106567  & $-0$&000120868 & $-5$&99954$\times10^{-5}$ \\
14  &    8&55174$\times10^{-5}$ & $-0$&000146607 
	& $-5$&15038$\times10^{-5}$ \\
15  &    7&61065$\times10^{-5}$ & $-9$&24268$\times10^{-5}$ 
	& $-4$&50976$\times10^{-5}$
\end{tabular}
\end{table}

In the massless limit there are several checks in the literature. 
As $M_1a\to 0$, we find
\begin{equation}
	Z_{M_1}^{[1]} = C_F\left[0.10726(15) + 0.04901(2) c_{\text{SW}}
		- 0.008735(5) c_{\text{SW}}^2 
		- \frac{3}{16\pi^2}\ln(M_1^{[0]}a)^2 \right].
\end{equation}
Note the appearance of the logarithm, multiplied by the 
one-loop anomalous dimension.
The finite part agrees well with published values: 
$C_F0.107347(5)$ at $c_{\text{SW}}=0$~\cite{GHS84}, and 
$C_F0.1474(4)$ at $c_{\text{SW}}=1$~\cite{Gab91}.
The individual coefficients of~$c_{\text{SW}}$ agree with 
Ref.~\cite{Cap97}.

Recently Sint and Weisz~\cite{Sin97} have computed the next term in 
the expansion of $Z_{M_1}^{[1]}(M_1a)$ around $M_1a=0$, 
for~$c_{\text{SW}}=1$.
They find the coefficient of~$M_1a$ to be $-C_F0.07217(2)$.
Fitting our results for $M_1a<0.1$, we find $-C_F0.0720(7)$, less 
precise, but in agreement.
Reference~\cite{Sin97} does not report a contribution of order
$M_1a\ln M_1a$; for $c_{\text{SW}}=1$ one would expect it to drop out.
Our Pauli-Villars subtractions isolate from~$Z_{M_1}^{[1](0)}$ a 
contribution $C_F[6/16\pi^2]M_1a\ln M_1a$, and our fits find nothing 
more of order $M_1a\ln M_1a$ in~$Z_{M_1}^{[1](0)}$.
But~$Z_{M_1}^{[1](1)}$ contains precisely the same contribution 
with the opposite sign, so the total drops out when $c_{\text SW}=1$.
This exercise verifies, as in Refs.~\cite{Gab91,Cap97}, that the 
tree-level improved action removes terms of order~$a\ln a$.

These checks are reassuring, but the main result is the full mass 
dependence, embodied in Figs.~\ref{fig:M1-1} and~\ref{fig:ZM1-1} and 
in Table~\ref{tbl:zM1-Cheb}.
To proceed from our numerical results to~$Z_{M_1}^{[1]}$:
\begin{enumerate}
	\item reconstitute adequate approximations to 
	the~$z_{M_1}^{[1](n)}$ from Table~\ref{tbl:zM1-Cheb};
	\item evaluate them at the desired value of $\tanh M_1^{[0]}$;
	\item accumulate the polynomial in $c_{\text{SW}}$;
	\item add
$e^{-M_1^{[0]}}\cosh M_1^{[0]}(p_{A_0}A_{\text{PV}}-p_CB_{\text{PV}})$ 
	to restore the Pauli-Villars subtraction.
\end{enumerate}
The full one-loop approximation to the rest mass is then
\begin{equation}
M_1 = M_1^{[0]} + g^2Z_{M_1}^{[1]} \tanh M_1^{[0]}.
\end{equation}
In a straightforward application of bare perturbation theory the 
expansion parameter~$g^2$ would be the bare coupling~$g_0^2$.
It is possible, however, to choose a better expansion 
parameter~\cite{Lep93}.
Further discussion of this issue will appear elsewhere.

\subsection{Kinetic Mass $M_2$}
\label{one-loop M2}
\begin{figure}[btp]
\begin{center}
	\epsfxsize=0.45454545\textwidth
	\epsfbox{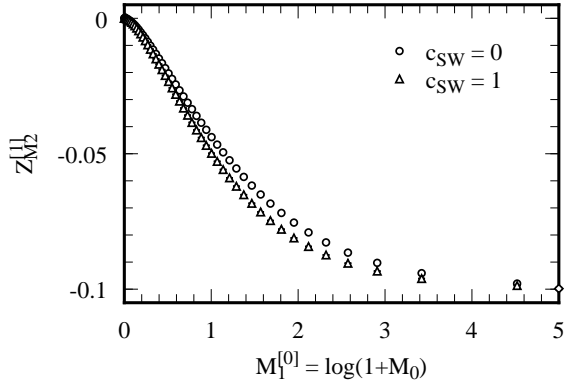}
\end{center}
\caption[fig:ZM2]{Plot of $Z_{M_2}^{[1]}$ vs.\ $M_1^{[0]}a$ for 
$c_{\text{SW}}=0$~(circles) and 1~(triangles).
The diamond marks the static limit~\cite{Dav92}.
Here $C_F=4/3$.}\label{fig:ZM2}
\end{figure}

\begin{figure}[btp]
\begin{center}
	\epsfxsize=0.45454545\textwidth
	\epsfbox{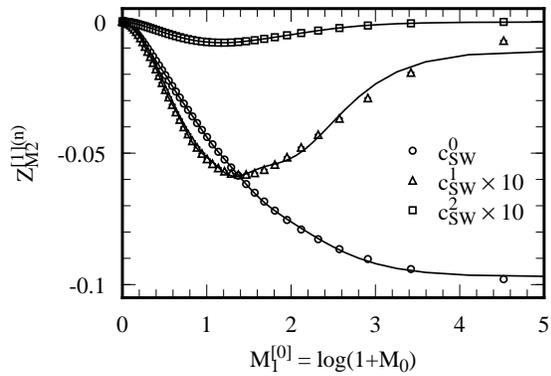}
\end{center}
\caption[fig:ZM2cSW]{Plot of $Z_{M_2}^{[1](n)}$ vs.\ $M_1^{[0]}a$.
The curves indicate the fifteen-term Chebyshev interpolation.
Note that the coefficients of $c_{\text{SW}}^1$ and $c_{\text{SW}}^2$ 
are multiplied by ten.  Here $C_F=4/3$.}\label{fig:ZM2cSW}
\end{figure}

Figure~\ref{fig:ZM2} shows the one-loop renormalization of the kinetic
mass for $c_{\text{SW}}=0$ and~1.
(The variation with~$c_{\text{SW}}$ is too weak to distinguish 1.4 
from~1.)
Again, $Z_{M_2}^{[1]}$ smoothly connects to the massless and static 
limits.
The separate coefficients of $c_{\text{SW}}$ are plotted in 
Fig.~\ref{fig:ZM2cSW}, and their first fifteen Chebyshev coefficients 
are listed in Table~\ref{tbl:ZM2-Cheb}.
\begin{table}[tbp]
\centering
\caption[tbl:ZM2-Cheb]{Coefficients of Chebyshev polynomials~$T_j(x)$ 
for~$Z_{M_2}^{[1](n)}$.  Here $C_F=4/3$.}\label{tbl:ZM2-Cheb}
\begin{tabular}{c*{3}{r@{.}l}}
$j$	& \multicolumn{2}{c}{$Z_{M_2}^{[1](0)}$}
	& \multicolumn{2}{c}{$Z_{M_2}^{[1](1)}$}
	& \multicolumn{2}{c}{$Z_{M_2}^{[1](2)}$} \\ 
\hline
 0  & $-$0&0631437   & $-$0&00518218  & $-$0&000697017 \\
 1  & $-$0&0409363   & $-$0&00266278  & $-$0&000293363 \\
 2  & $-$0&0113513   &    0&000649031 &    0&000188227 \\
 3  & $-$0&00411835  &    0&00104410  &    0&000188092 \\
 4  & $-$0&00293037  &    0&000578677 &    9&36398$\times10^{-5}$ \\
 5  & $-$0&00157589  &    0&000435851 &    5&85129$\times10^{-5}$ \\
 6  & $-$0&00118956  &    0&000331139 &    3&40842$\times10^{-5}$ \\
 7  & $-$0&00081670  &    0&000251324 &    2&23707$\times10^{-5}$ \\
 8  & $-$0&000630783 &    0&000204140 &    1&47425$\times10^{-5}$ \\
 9  & $-$0&000490657 &    0&000162222 &    1&02279$\times10^{-5}$ \\
10  & $-$0&000396438 &    0&000137218 &    0&71930$\times10^{-5}$ \\
11  & $-$0&000319830 &    0&000112310 &    0&52123$\times10^{-5}$ \\
12  & $-$0&000273443 &    9&74213$\times10^{-5}$ 
    &    0&39003$\times10^{-5}$ \\
13  & $-$0&000223656 &    8&30367$\times10^{-5}$ 
    &    0&29204$\times10^{-5}$ \\
14  & $-$0&000198908 &    7&12494$\times10^{-5}$ 
    &    0&21814$\times10^{-5}$ \\
15  & $-$0&000163911 &    6&31163$\times10^{-5}$ 
    &    0&17610$\times10^{-5}$
\end{tabular}
\end{table}

In the static limit we find 
$Z_{M_2}^{[1]}(\infty)=-C_F0.0745(1)=-0.0993(1)$, 
which agrees with $-0.0998(4)$, the value in nonrelativistic 
QCD~\cite{Dav92,Mor93} for our definition of~$Z_{M_2}$.
In the massless limit we verify $Z_{M_2}^{[1]}(0)=0$, because there 
$M_2=M_1$.

These checks are again reassuring, but the main result is the full 
mass dependence, embodied in Figs.~\ref{fig:ZM2} and~\ref{fig:ZM2cSW} 
and in Table~\ref{tbl:ZM2-Cheb}.
To proceed from our numerical results to~$Z_{M_2}^{[1]}$:
\begin{enumerate}
	\item reconstitute adequate approximations to 
	the~$Z_{M_2}^{[1](n)}$ from Table~\ref{tbl:ZM2-Cheb};
	\item evaluate them at the desired $\tanh M_1^{[0]}$;
	\item accumulate the polynomial in $c_{\text{SW}}$.
\end{enumerate}
The full one-loop approximation to the kinetic mass is then
\begin{equation}
M_2=m_2(M_1^{[0]}+g_1^2M_1^{[1]})\left(1 + g_2^2Z_{M_2}^{[1]}\right).
\label{M2 one loop}
\end{equation}
Again, it may be appropriate to choose optimal expansion 
parameters~$g_1^2$ and~$g_2^2$, as will be discussed elsewhere.
 
\subsection{Wave-function Renormalization $Z_2$}
\label{one-loop Z2}
\begin{figure}[btp]
\begin{center}
	\epsfxsize=0.45454545\textwidth
	\epsfbox{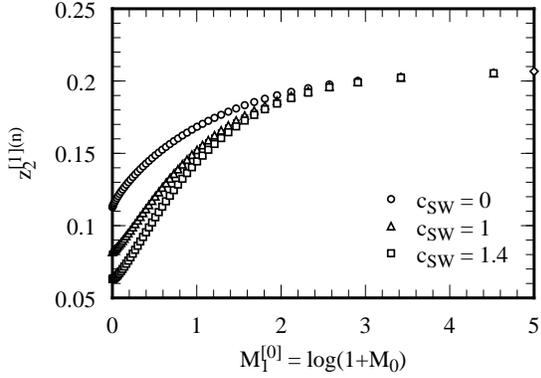}
\end{center}
\caption[Z2-1]{Plot of $z_2^{[1]}$ vs.\ $M_1^{[0]}$ for
$c_{\text{SW}}=0$~(circles), 1~(triangles), and 1.4~(squares).
The diamond marks the static limit~\cite{Eic90}.
Here $C_F=4/3$.}\label{fig:Z2}
\end{figure}

\begin{figure}[btp]
\begin{center}
	\epsfxsize=0.45454545\textwidth
	\epsfbox{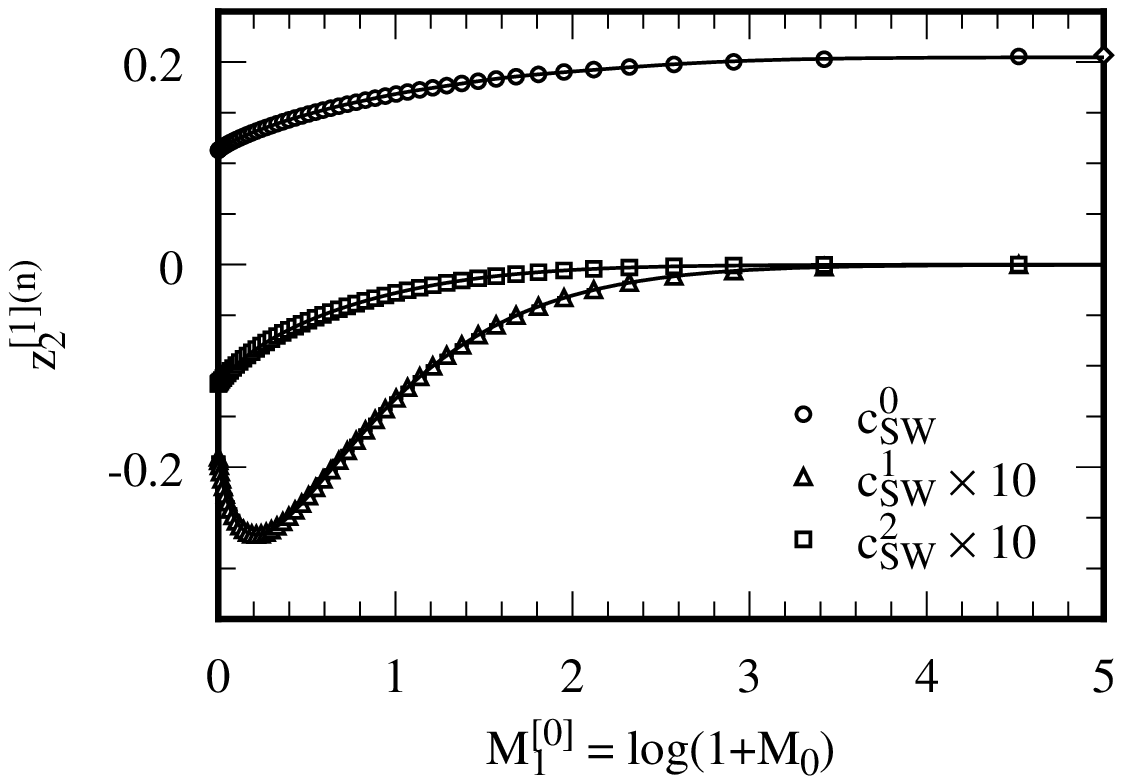}
\end{center}
\caption[fig:Z2cSW]{Plot of $z_2^{[1](n)}$ vs.\ $M_1^{[0]}a$.
The curves indicate the fifteen-term Chebyshev interpolation.
Note that the coefficients of $c_{\text{SW}}^1$ and $c_{\text{SW}}^2$ 
are multiplied by ten.  Here $C_F=4/3$.}\label{fig:Z2cSW}
\end{figure}

Because the wave-function renormalization factor's full 
correction~$Z_2^{[1]}$ has an infrared divergence for all values of 
the quark mass, we present results for the subtracted form, as defined 
in Eq.~(\ref{Z2 sub}).
Figure~\ref{fig:Z2} shows the one-loop correction~$z_2^{[1]}$, in 
Feynman gauge, for $c_{\text{SW}}=0$, 1, and~1.4.
Once again, $z_2^{[1]}$ smoothly connects the massless and static 
limits.
The separate coefficients of $c_{\text{SW}}$ are plotted in 
Fig.~\ref{fig:Z2cSW}, and their first fifteen Chebyshev coefficients 
are listed in Table~\ref{tbl:Z2-Cheb}.
\begin{table}[tbp]
\centering
\caption[tbl:Z2-Cheb]{Coefficients of Chebyshev polynomials~$T_j(x)$ 
for~$z_2^{[1](n)}$.  Here $C_F=4/3$.} \label{tbl:Z2-Cheb}
\begin{tabular}{c*{3}{r@{.}l}}
$j$	& \multicolumn{2}{c}{$z_2^{[1](0)}$}
	& \multicolumn{2}{c}{$z_2^{[1](1)}$}
	& \multicolumn{2}{c}{$z_2^{[1](2)}$} \\ 
\hline
 0  & 0&305389    & $-$0&033722    & $-$0&0111230   \\
 1  & 0&0392549   &    0&0108371   &    0&00554548  \\
 2  & 0&00310868  &    0&00597098  & $-$0&000324921 \\
 3  & 0&00393177  & $-$0&00107846  &    0&000324902 \\
 4  & 0&00138399  &    0&00096398  & $-$1&79973$\times10^{-5}$ \\
 5  & 0&00134333  & $-$0&000166528 &    2&44776$\times10^{-5}$ \\
 6  & 0&000666205 &    0&000233668 &    0&38057$\times10^{-5}$ \\
 7  & 0&000655838 & $-$4&30696$\times10^{-5}$ 
    &    0&283444$\times10^{-5}$ \\
 8  & 0&000391408 &    8&67030$\times10^{-5}$ 
    &    0&156463$\times10^{-5}$ \\
 9  & 0&000380178 & $-$1&79973$\times10^{-5}$ 
    &    0&062247$\times10^{-5}$  \\
10  & 0&000255127 &    4&08834$\times10^{-5}$ 
    &    0&038319$\times10^{-5}$  \\
11  & 0&000252910 & $-$1&04220$\times10^{-5}$ 
    &    0&020393$\times10^{-5}$  \\
12  & 0&000181397 &    2&24672$\times10^{-5}$ 
    &    0&007974$\times10^{-5}$  \\
13  & 0&000172181 & $-$0&68263$\times10^{-5}$ 
    &    0&005127$\times10^{-5}$  \\
14  & 0&000122340 &    1&43152$\times10^{-5}$ 
    &    0&001010$\times10^{-5}$  \\
15  & 0&000120061 & $-$0&53572$\times10^{-5}$ 
    & $-$0&004476$\times10^{-5}$
\end{tabular}
\end{table}

In the static limit we find
$Z_2^{[1]}(\infty)+C_F[2/16\pi^2]\ln\lambda^2=C_F0.1548(5)=0.2064(7)$, 
which agrees with $0.2067(1)$, the value of Refs.~\cite{Eic90,Dav92} 
for our definition of~$Z_2$.

In the massless limit we find 
\begin{eqnarray}
	Z_2^{[1]}
		&=&	C_F\left[0.05608(7) - 0.014239(6) c_{\text{SW}}
		- 0.008844(3) c_{\text{SW}}^2 \right. \label{Z2 massless} \\
		&+&	\left.
		\left(3\ln(M_1^{[0]}a)^2 -2\ln(\lambda a)^2\right)/16\pi^2
		\right]. \nonumber
\end{eqnarray}
Note the appearance of a logarithm of~$M^2$ as well as the infrared 
divergence.
The finite part agrees well with published values: 
$C_F0.056057(2)$ at $c_{\text{SW}}=0$~\cite{GHS84} and 
$C_F0.0329(3)$   at $c_{\text{SW}}=1$~\cite{Gab91}.
The individual coefficients of~$c_{\text{SW}}$ agree with 
Ref.~\cite{Cap97}.
For comparison with Figs.~\ref{fig:Z2} and~\ref{fig:Z2cSW}, note 
that though~$z_2^{[1](0)}$ does not contain the logarithms, it is 
larger by~$C_F[9/32\pi^2]$ than the finite part of~$Z_2^{[1](0)}$.

Once again, the checks are reassuring, but the main result is the full 
mass dependence.
In practice, the wave-function renormalization factor is used 
only combined with vertex renormalization factors, in ways such 
that the infrared divergences cancel.
When calculating a vertex renormalization factor, one should isolate 
all infrared divergences analytically, as we have done here, and then 
assemble the pieces so that the cancellation is explicit.
If one chooses an on-shell renormalization scheme, dependence on the 
gauge parameter should cancel as well.

\section{Improved Perturbation Theory}
\label{tadpoles removed}

The previous section presented results for the (mass-dependent) 
one-loop coefficients in Eqs.~(\ref{M1 series}), (\ref{ZM2 series}), 
and~(\ref{Z2 series}).
Truncating the series at one loop, without further ado, is unlikely 
to be a good approximation to the full series, however.
The one-loop self energy is large, owing to the contribution of the 
tadpole diagram in Feynman gauge,~$\Sigma^{[1]}_{\text{(b)}}$.
This feature will persist at higher orders.
Similarly, tadpole diagrams make the bare coupling~$g_0^2$ much 
smaller than other, more physical measures of the gauge interaction.
Thus, one must be wary of these series; they are characterized by an 
unusually small expansion parameter, counteracted by unusually large 
coefficients.

These observations suggest rearranging perturbation series so that 
tadpole diagrams cancel each other in final results~\cite{Lep93}.
The rearrangement is achieved by defining new couplings and 
(re)normalization factors.
Wherever the gauge field appears in the action or in operators, 
substitute
\begin{equation}
	U_\mu\to u_0[U_\mu/u_0],
	\label{u0}
\end{equation}
where the new parameter~$u_0$, the mean link, is a gauge-invariant 
average of the link matrices~$U_\mu$.
The perturbative expansion of~$u_0$, like any involving~$U_\mu$, is 
dominated by its tadpole diagram.
In the perturbative expansion of~$U_\mu/u_0$, on the other hand, the  
tadpole diagrams cancel.
The price for arranging the cancellation is that the first factor 
of~$u_0$ in Eq.~(\ref{u0}) must be evaluated nonperturbatively.

In the following discussion we shall denote the tadpole-improved 
gauge coupling as~$\tilde{g}^2$, without defining it explicitly.
The definition of~$\tilde{g}^2$ affects higher-order corrections and 
numerical evaluation of a series truncated at one loop (or beyond).
But it does not affect the one-loop self energy, which is the main 
focus here.
On the other hand, tadpole improvement of the bare mass and of the
field normalization do alter the one-loop coefficients, and they are 
considered below.
 
\subsection{Rest Mass $M_1$}
\label{tadpole one-loop M1}
The rest mass is primarily sensitive to the bare mass, particularly 
the subtracted version, $M_0=m_0-m_{0c}$.
To derive a tadpole-improved bare mass, it is easiest to apply 
Eq.~(\ref{u0}) to the hopping-parameter form of the action.
Our action $S_0$~\cite{KKM97} has two hopping parameters,
\begin{equation}
	\kappa_t = \frac{1}{2[1 + (d-1)r_s\zeta + m_0]}
	\label{kappa t}
\end{equation}
for temporal hops and $\kappa_s=\zeta\kappa_t$ for spatial hops.
After rescaling the fields by~$\sqrt{2\kappa_t}$, the link matrices 
appear only in the combinations~$\kappa_t U_0$ and~$\kappa_s U_i$.
In view of Eq.~(\ref{u0}) one defines the tadpole-improved hopping 
parameters $\tilde{\kappa}_{t,s}=u_0\kappa_{t,s}$ to absorb the first 
factor of~$u_0$.
Undoing the rescaling, now with~$\sqrt{2\tilde{\kappa}_t}$, and 
reassembling the mass form leads to
\begin{equation}
\tilde{M}_0 \equiv \tilde{m}_0-\tilde{m}_{0c} = M_0/u_0.
\end{equation}
The numerical value of the tadpole-improved mass~$\tilde{M}_0$ relies 
on two nonperturbatively determined parameters, $m_{0c}$ and~$u_0$.

This rearrangement does not alter Eq.~(\ref{M1M0}).
Before developing the perturbation series, however, one sets 
$M_0=u_0\tilde{M}_0$, treats~$\tilde{M}_0$ as the variable independent 
of~$\tilde{g}^2$, and expands the explicit~$u_0$ as well as the 
self-energy functions.
Thus,
\begin{equation}
	M_1 = \sum_{l=0}^\infty \tilde{g}^{2l} \tilde{M}_1^{[l]},
	\label{M1 tadpole series}
\end{equation}
where $\tilde{M}_1^{[0]}=\ln(1+\tilde{M}_0)$ and
\begin{equation}
	\tilde{M}_1^{[1]} = M_1^{[1]} + 
	\frac{\tilde{M}_0}{1+\tilde{M}_0}u_0^{[1]}.
	\label{tilde M1[1]}
\end{equation}
Since $u_0^{[1]}$ is negative, the (positive) coefficient~$M_1^{[1]}$ 
is reduced.
Note also that the mass dependence of the tadpole improvement 
connects smoothly to the small- and large-mass limits.

\begin{figure}[btp]
\begin{center}
	\epsfxsize=0.45454545\textwidth
	\epsfbox{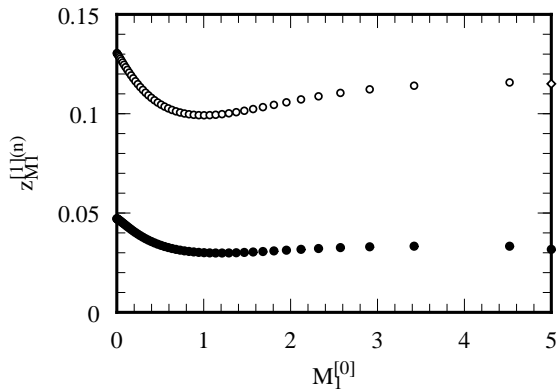}
\end{center}
\caption[fig:ZM1tad]{Plot comparing $z_{M_1}^{[1]}a$ vs.\ $M_1^{[1]}a$ 
(open symbols) and $\tilde{z}_{M_1}^{[1]}a$ vs.\ $\tilde{M}_1^{[1]}a$ 
(closed symbols).  The static limit is indicated by 
diamonds~\cite{Eic90}.}\label{fig:ZM1tad}
\end{figure}

To illustrate the efficacy of tadpole improvement we take the mean 
link to be $u_0=\langle U_{\Box}\rangle^{1/4}$, where~$U_\Box$ denotes 
the product of link matrices around a plaquette, $u_0^{[1]}=-C_F/16$.
Figure~\ref{fig:ZM1tad} shows explicitly 
that~$\tilde{z}_{M_1}^{[1]}$ is significantly smaller  
than~$z_{M_1}^{[1]}$, when~$u_0$ is defined this way.
(This choice of~$u_0$ does not modify the terms proportional 
to~$c_{\text{SW}}$ and~$c_{\text{SW}}^2$.)
For $c_{\text{SW}}=1$ the mass dependence at $Ma=0$ is also less steep: 
$-C_F[0.07217-1/32]=-C_F0.04088$.
Other definitions of~$u_0$ produce a qualitatively similar reduction.

\subsection{Kinetic Mass $M_2$}
\label{tadpole one-loop M2}

Because the kinetic-mass renormalization factor~$Z_{M_2}$ is defined 
as a ratio of on-shell (and therefore physical) quantities, it should 
not be surprising that its one-loop correction~$Z_{M_2}^{[1]}$ is 
unaffected by the tadpole diagram.
More concretely, one can track analytically the contribution 
of~$\Sigma^{[1]}_{(b)}$ through Eq.~(\ref{ZM2[1]}), to verify that it 
drops out (for all~$M_1^{[0]}$).
Thus, there are only two changes to improve the one-loop estimate of 
the kinetic mass:
Use $\tilde{M}_1^{[0]}+\tilde{g}^2\tilde{M}_1^{[1]}$ as 
the argument of~$m_2$ in Eq.~(\ref{M2 one loop}), and
use an improved coupling in the expansion of~$Z_{M_2}$.

\subsection{Wave-function Renormalization $Z_2$}
\label{tadpole one-loop Z2}
\begin{figure}[btp]
\begin{center}
	\epsfxsize=0.45454545\textwidth
	\epsfbox{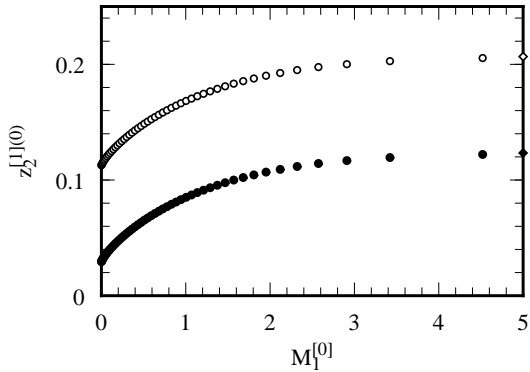}
\end{center}
\caption[Z2-1]{Plot comparing $z_2^{[1]}$ vs.\ $M_1^{[1]}a$ (open 
symbols) and $\tilde{z}_2^{[1]}$ vs.\ $\tilde{M}_1^{[1]}a$ (closed
symbols).}\label{fig:Z2tad}
\end{figure}

In the rescaling from the hopping-parameter form of the action to the 
mass form, the tadpole-improved version carries an additional factor 
of~$u_0$, because $\tilde{\kappa}=u_0\kappa$.
Thus, the renormalization factor of the tadpole-improved wave function 
is $\tilde{Z}_2=u_0Z_2$.
Following Eq.~(\ref{Z2 series}) we develop the perturbative series
\begin{equation}\label{Z2 tadpole series}
e^{M_1a}\tilde{Z}_2=
1+\sum_{l=1}^\infty \tilde{g}^{2l}\tilde{Z}_2^{[l]}.
\end{equation}
Comparing the expansions one easily finds
\begin{equation}
\tilde{Z}_2^{[1]}=Z_2^{[1]}+u_0^{[1]}.
\end{equation}
Figure~\ref{fig:Z2tad} compares 
$\tilde{z}_2^{[1]}=z_2^{[1]}+u_0^{[1]}$ to~$z_2^{[1]}$
(in Feynman gauge).
Once again, $\tilde{z}_2^{[1]}$ is significantly smaller.

\section{Conclusions}
\label{conclusions}
The work presented here is the first thorough study of renormalization 
in the approach of Ref.~\cite{KKM97} to massive lattice fermions.
(Some of our results have appeared earlier~\cite{Mer94,Mer95,Mer97}.)
On the one hand, the specific, numerical results can be combined with 
Monte Carlo calculations of heavy-quark spectra to determine the heavy 
quarks' masses.
On the other hand, the techniques used to arrive at those results can 
be extended as the renormalization program is extended to vertex 
corrections.
The most timely example of the latter is the renormalization of 
(improved) vector and axial-vector currents~\cite{Aok98}, which are
needed to obtain heavy-light meson decay constants and semi-leptonic 
form factors.

The theoretical analysis of Sec.~\ref{quark pole} examines 
renormalization to all orders in the gauge coupling and in the fermion 
mass.
Although the focus is on the fermion propagator, the analysis serves 
as a model for the renormalization of currents as well.
For currents one would introduce appropriate vertex functions, the 
analogs of the self-energy functions~$A_\rho$ and~$C$, again 
constrained only by periodicity and symmetry properties.
Then one would Fourier transform the fourth component of each external 
momentum to obtain the on-shell correlation function (for quark 
states) to all orders in the mass and the gauge coupling.
Just as here, the all-orders formulae would provide useful insights, 
and they would be convenient for developing expansions to any desired 
order.

In our numerical work we subtract from the self-energy functions 
corresponding functions from a Pauli-Villars regulator.
In this way we are able to isolate mass and infrared singularities,
including, for small~$Ma$, contributions of the form $Ma\ln Ma$.
The Pauli-Villars functions can be written in closed form, so we have 
analytical control of the singularities.
Furthermore, by applying the subtraction at the integrand level, 
numerical integration of the nonsingular part is simplified.
For example, one can carry out the integration without an infrared 
regulator.
These techniques should continue to be helpful for the renormalization 
of currents and other operators.

The numerical results presented in Sec.~\ref{results} demonstrates 
that the mass dependence of renormalization factors smoothly connects 
the massless and static limits.
This was expected from the free theory and general 
arguments~\cite{KKM97}, but it is satisfying to make it explicit.
Also expected, but now explicit, is the result in 
Sec.~\ref{tadpoles removed} that tadpole improvement~\cite{Lep93} 
reduces the size of the one-loop coefficients for all masses.
In the case of the rest-mass renormalization factor, tadpole 
improvement also makes the mass dependence even smoother, cf.\ 
Fig.~\ref{fig:ZM1tad}.

Our results, especially with tadpole improvement, will be useful in 
determining the $\overline{\rm MS}$ masses of heavy 
quarks~\cite{Kro97}.
Additionally, such a determination will require at least an evaluation 
of the optimal scale~\cite{Lep93} (in progress), and, to achieve 
better than 5\% accuracy, the complete two-loop calculation.
Nevertheless, the approach taken here to the renormalization of 
massive lattice gauge theories is a necessary first step.

\section*{Acknowledgments}
We are grateful to Paul Mackenzie for many helpful discussions.
We thank Carlotta Pittori for correspondence on Ref.~\cite{Gab91},
and Stefano Capitani for correspondence on Ref.~\cite{Cap97}.
B.P.G.M. is supported in part by the U.S. Department of Energy under
Grant No.\ DE-FG02-90ER40560.
A.X.K. is supported in part by the OJI program of the U.S.  
Department of Energy under Grant No.\ DE-FG02-91ER40677 and by a 
fellowship from the Alfred P. Sloan Foundation.
Fermilab is operated by Universities Research Association, Inc.,
under contract with the U.S. Department of Energy.

\appendix \widetext
\section{Feynman Rules}
\label{feynman rules}

\begin{figure}
  \begin{center}
  \epsfbox{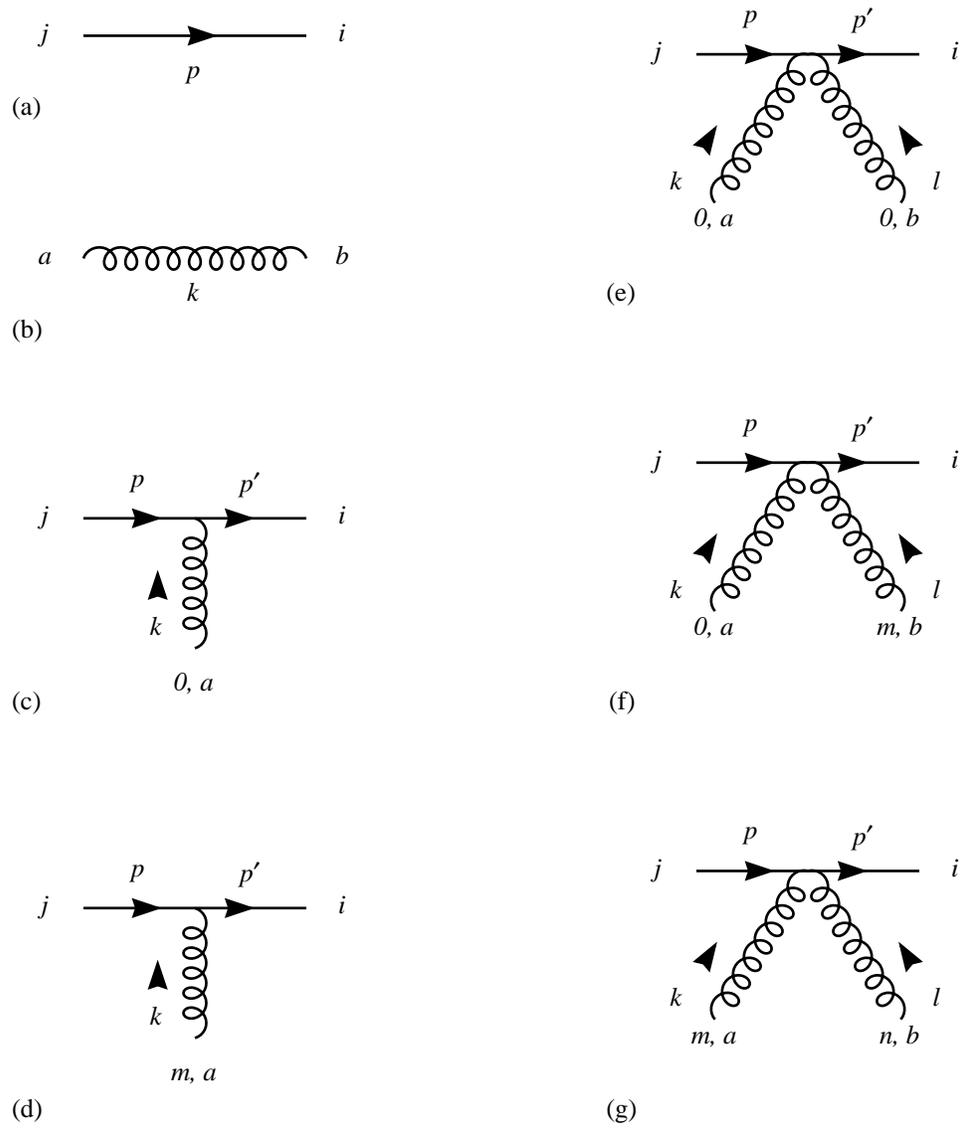}
  \end{center}
  \caption[rules]{Feynman rules for the action $S_0+S_B+S_E$ defined 
  in Sec.~\ref{action}.}\label{rules}
\end{figure}

This is the first paper to address perturbation theory for the action 
discussed in Sec.~\ref{action}, so we give here the propagators and 
vertices needed for the one-loop self energy.
Figure~\ref{rules} defines indices and momentum flow.
Then
\begin{eqnarray}
\text{Fig.~\ref{rules}a} &=& G_0(p)_{ij}
	= \frac{\delta_{ij}}%
	{i\gamma_0\sin p_0 + i\zeta\bbox{\gamma}\cdot\bbox{S}(\bbox{p})
	+ 1 + m_0 - \cos p_0 + \case{1}{2} r_s \zeta \hat{\bbox{p}}^2},
	\\[1.0em]
\text{Fig.~\ref{rules}b} &=& \Delta^{ab}_{\mu\nu}(k)
	= \frac{\delta^{ab}}{\hat{k}^2} \left( \delta_{\mu\nu}
	- (1-\alpha)\frac{\hat{k}_{\mu} \hat{k}_{\nu}}{\hat{k}^2} \right),
	\label{gluon prop} \\[1.0em]
\text{Fig.~\ref{rules}c} &=& - g_0  t^a_{ij} [
	\gamma_0 \cos\case{1}{2}(p+p')_0 - i \sin\case{1}{2} (p+p')_0 
  +	\case{1}{2} c_E \zeta \sigma_{0r} \cos\case{1}{2}k_0 \sin k_r ],
	\\[1.0em]
%
\text{Fig.~\ref{rules}d} &=& - g_0 \zeta t^a_{ij} [ 
	\gamma_m \cos\case{1}{2}{(p+p')_m} - i r_s \sin\case{1}{2}(p+p')_m
	\\ \nonumber
    &+& \case{1}{2} c_B \sigma_{mn} \cos\case{1}{2}k_m \sin k_n
    + \case{1}{2} c_E \sigma_{m0} \cos\case{1}{2}k_m \sin k_0 ],
     \\[1.0em]
\text{Fig.~\ref{rules}e} &=& g_0^2 \case{1}{2}\{t^a,t^b\}_{ij} [
	\cos\case{1}{2}(p+p')_0 - i \gamma_0 \sin\case{1}{2}(p+p')_0 ]
	\\ \nonumber
	&-&	\case{i}{4} g_0^2 c_E\zeta [t^a,t^b]_{ij}
	\sigma_{0r} (\sin k_r-\sin l_r) \sin\case{1}{2}(k+l)_0,
	\\[1.0em]
\text{Fig.~\ref{rules}f} &=&-\case{i}{2} g_0^2 c_E\zeta 
	[t^a,t^b]_{ij}\sigma_{0m} \times \\ & & \nonumber
   [2 \cos\case{1}{2}(k+l)_0 \cos\case{1}{2}(k+l)_m
      \cos\case{1}{2}l_0 \cos\case{1}{2}k_m 
    - \cos\case{1}{2}l_m \cos\case{1}{2}k_0 ], \\[1.0em]
\text{Fig.~\ref{rules}g}
  &=& \case{1}{2} g_0^2  \zeta   \{t^a,t^b\}_{ij} \delta_{mn}  
  [r_s \cos\case{1}{2}(p+p')_m - i\gamma_m \sin\case{1}{2}(p+p')_m] \\
  &-& \case{i}{2} g_0^2 c_B\zeta  [t^a,t^b]_{ij}  \sigma_{mn} 
  	 \times \nonumber \\ & &
    [2\cos\case{1}{2}(k+l)_m \cos\case{1}{2}(k+l)_n
      \cos\case{1}{2}l_m \cos\case{1}{2}k_n
   -  \cos \case{1}{2}l_n \cos \case{1}{2}k_m ] \nonumber \\
  &+& \case{i}{4} g_0^2 \zeta [t^a,t^b]_{ij}
    \delta_{mn} \sin\case{1}{2}(k+l)_m
    [ c_B \sigma_{mr} (\sin k_r - \sin l_r)
    + c_E \sigma_{m0} (\sin k_0 - \sin l_0) ], \nonumber
\end{eqnarray}
where $S_i(p_i)=\sin p_i$;
$\alpha$ is the gauge parameter ($\alpha=1$ yields Feynman gauge);
$g_0$~is the bare gauge coupling;
the $t^a$~are (anti-Hermitian) generators of SU(3), normalized so that 
$\mathop{\rm tr}t^at^b=-\delta^{ab}/2$;
and the $\gamma$ matrix conventions are as in Ref.~\cite{KKM97}.
One can verify easily that these rules reduce to the ones for the
Sheikholeslami-Wohlert action when $\zeta=1$, $r_s=1$, and $c_B=c_E$.

\section{Numerators}
\label{numerators}
It is convenient to sort the numerator~$F(p,k)$ in Eq.~(\ref{Sigma 1}) 
according to the number of clover interactions:
\begin{equation}
	F = F_0 + c_B F_{c_B} + c_E F_{c_E}
		+ c_B^2 F_{c_B^2} + c_B c_E F_{c_Bc_E} + c_E^2 F_{c_E^2}.
	\label{FFF}
\end{equation}
The Wilson-action term~$F_0$ is gauge dependent, but the other terms 
are independent of the gauge parameter.
Below we give~$F_0$ in Feynman gauge [$\alpha=1$ in 
Eq.~(\ref{gluon prop})].
The sum of internal quark and gluon momenta appears often, so let 
$s=p+(p+k)=2p+k$.
From the Feynman rules, tedious algebra (performed by Mathematica and 
checked by hand) yields
\begin{eqnarray}
F_0(p,k)
	&=& i\gamma_0 \left[ \sin(p+k)_0 \{\cos s_0 
	 -	\zeta^2 [d-1 + \case{1}{4}(r_s^2 -1)\hat{\bbox{s}}^2 ]\}
	 +	L(p+k) \sin s_0 \right] \nonumber \\  
	&+& i\zeta \sum_i \gamma_i \left[
		\sin(p+k)_i \left(\zeta^2 \{\cos s_i 
	 -	[d-1 + \case{1}{4}(r_s^2 -1)\hat{\bbox{s}}^2]\}
	 +	\zeta^2 - 1 \right) \right. \nonumber \\
	& &	\left. \hspace{2.5cm} \rule{0pt}{1em}
	  +	r_s \zeta L(p+k) \sin s_i \right] \nonumber \\
	&+& \sin s_0 \sin(p+k)_0
	 +	r_s \zeta^3 \sin\bbox{s} \cdot \sin(\bbox{p}+\bbox{k}) 
	 \nonumber \\
	&-&	L(p+k)\{\cos s_0
	 +	\zeta^2[d-1 - \case{1}{4}(r_s^2 +1)\hat{\bbox{s}}^2]\},
	 \label{F0} 
\end{eqnarray}
where $\hat{\bbox{q}}^2=\sum_i(2\sin\case{1}{2}q_i)^2$,
and $L(q)=1+m_0a+\case{1}{2}r_s\zeta\hat{\bbox{q}}^2-\cos q_0$;
\begin{eqnarray}
F_{c_B}
	&=& ir_s \zeta^3 \sum_{j\neq i}
		[\gamma_i \sin(p+k)_j - \gamma_j \sin(p+k)_i]
	 	\sin k_i \sin\case{1}{2}s_j  \cos\case{1}{2}k_j \nonumber \\
	&-& i\zeta^2 L(p+k)\sum _{j\neq i} \gamma_i \sin k_i
		\cos\case{1}{2}k_j \cos\case{1}{2}s_j 
	+	\zeta^3 \sum_{j\neq i} \sin k_i \sin(p+k)_i
		\cos\case{1}{2}k_j \cos\case{1}{2}s_j;		
\end{eqnarray}
\begin{eqnarray}
F_{c_E}
	&=& ir_s \zeta^3 \gamma_0 \sin k_0 \sum_i \cos\case{1}{2}k_i 
		\sin\case{1}{2}s_i \sin(p+k)_i \nonumber \\
	&-& i\zeta^2 \gamma_0 \left[\sin\case{1}{2}s_0 \cos\case{1}{2}k_0
		\sin\bbox{k}\cdot\sin(\bbox{p}+\bbox{k})
	 + 	\sin k_0 L(p+k) \sum_i \cos\case{1}{2}k_i \cos\case{1}{2}s_i 
	 	\right] \nonumber \\
	&+& i\zeta \bbox{\gamma}\cdot\sin\bbox{k} \cos\case{1}{2}k_0 
		[\sin(p+k)_0 \sin\case{1}{2}s_0 - L(p+k)\cos\case{1}{2}s_0] \nonumber \\
	&-& ir_s \zeta^2 \sin k_0 \sin(p+k)_0 
		\sum_i \gamma_i \cos\case{1}{2}k_i \sin\case{1}{2}s_i \nonumber \\
	&+& \zeta^2 \sin\bbox{k}\cdot\sin(\bbox{p}+\bbox{k})
		\cos\case{1}{2}k_0 \cos\case{1}{2}s_0 
	 +	\zeta^2 \sin k_0 \sin(p+k)_0
	 	\sum_i \cos\case{1}{2}k_i \cos\case{1}{2}s_i;
\end{eqnarray}
\begin{eqnarray}
F_{c_B^2}
	&=& -\case{i}{4}\zeta^2 \gamma_0 \sin (p+k)_0
		\sum_{j\neq i} \sin^2 k_i \cos^2\case{1}{2}k_j \nonumber \\
	&+& \case{i}{2}\zeta^3 \bbox{\gamma}\cdot\sin\bbox{k}
		\sum_{j\neq i} \sin k_i \sin(p+k)_i \cos^2\case{1}{2}k_j
	 -	\case{i}{4}\zeta^3 \bbox{\gamma}\cdot\sin(\bbox{p}+\bbox{k}) 
		\sum_{j\neq i} \sin^2 k_i \cos^2\case{1}{2}k_j \nonumber \\ 
	&+& \case{i}{2}\zeta^3 \sum_{{j\neq i}}
		\gamma_i \cos^2\case{1}{2}k_i \sin k_j
		[\sin(p+k)_i \sin k_j - \sin(p+k)_j \sin k_i] \nonumber \\
	&+& \case{1}{4}\zeta^2 L(p+k)
		\sum_{j\neq i} \sin^2 k_i \cos^2\case{1}{2}k_j ;
\end{eqnarray}
\begin{equation}
F_{c_Ec_B} =
	\case{i}{2}\zeta^3    \gamma_0 \sin k_0 
	\sum_{j\neq i} \sin(p+k)_i \sin k_i \cos^2\case{1}{2}k_j
 +	\case{i}{2}\zeta^2 \sin(p+k)_0 \sin k_0
	\sum_{j\neq i}    \gamma_i \sin k_i \cos^2\case{1}{2}k_j ;
\end{equation}
\begin{eqnarray}
F_{c_E^2}
	&=& \case{1}{4} \zeta^2 \left[i\gamma_0\sin(p+k)_0
	 -	i\zeta \bbox{\gamma}\cdot\sin(\bbox{p}+\bbox{k}) + L(p+k)
	 \right] \times \nonumber \\ & & \hspace{2.5cm}
	 \left[	\cos^2\case{1}{2}k_0 \sum_i \sin^2 k_i
		   +	\sin^2 k_0 \sum_i \cos^2\case{1}{2}k_i
	 \right] \nonumber \\
	&+& \case{i}{2}\zeta^3 \left[ (\bbox{\gamma}\cdot\sin\bbox{k})
		[\sin\bbox{k}\cdot\sin(\bbox{p}+\bbox{k})] \cos^2\case{1}{2}k_0
	 + \sin^2 k_0 \sum_i \gamma_i \sin(p+k)_i \cos^2\case{1}{2}k_i
	 \right]. \label{FcE2}
\end{eqnarray}

\narrowtext
\section{Integration over Loop Momentum}
\label{integration}

To apply Eqs.~(\ref{Z2}), (\ref{M1}), and~(\ref{M2}), one must obtain 
the self-energy functions~$A_\rho$ and~$C$ for real~$p_0$ and 
analytically continue them to imaginary values~$iE$.
The analytical continuation must be approached cautiously in the 
representation of the self-energy functions as integrals over the loop 
variable~$k_0$.
The subtleties arise even with the continuum self energy.
Setting the internal quark momentum to~$q$, one has
\begin{equation}
	\Sigma^{{[1]}}(p) = C_F
	\int\frac{d^dq}{(2\pi)^d}\,
	\frac{dm + i(d-2)\kern+0.1em /\kern-0.55em q}{(p-q)^2(q^2 + m^2)}.
	\label{continuum C}
\end{equation}
In the complex~$q_0$ plane there are poles at $\pm iE(\bbox{q})$ and 
at~$p_0\pm i|\bbox{p}-\bbox{q}|$.
(Here, in the continuum, $E(\bbox{q})=\sqrt{\bbox{q}^2+m^2}$.)
The integration contour is the real axis.
If one sets $p_0=iE(\bbox{p})$, however, pole at 
$p_0-i|\bbox{p}-\bbox{q}|$ crosses into the upper half-plane when 
$|\bbox{p}-\bbox{q}|<E(\bbox{p})$, and the contour must be deformed to 
accommodate it.
Numerical integration packages are not clever enough to deform the 
contour; it is usually the real axis by default.

A better choice is to let $q=p+k$.
In the complex~$k_0$ plane the poles are at 
$-p_0\pm iE(\bbox{p}+\bbox{k})$ and $\pm i|\bbox{k}|$.
Then setting $p_0=iE(\bbox{p})$ moves a pole across the real axis if 
$|\bbox{p}|>|\bbox{p}+\bbox{k}|$.
For the wave function and the rest mass, one wants external momentum 
$\bbox{p}=\bbox{0}$, so the pole cannot, in fact, cross.
When the quark mass is large, however, the pole comes very close to 
the real axis, producing a sharp peak near $k_0=0$.
Such integrals are difficult to estimate robustly.
Even worse are the derivatives needed for the combination~$D$, 
cf.~Eq.~(\ref{D}): partial differentiation with respect to $p_1$ and 
$p_0$ each induces infrared divergences, which must, however, cancel 
in the total derivative.

\begin{figure}[tbp]
	\begin{center}
	\epsfysize=0.80\textheight 
	\epsfbox{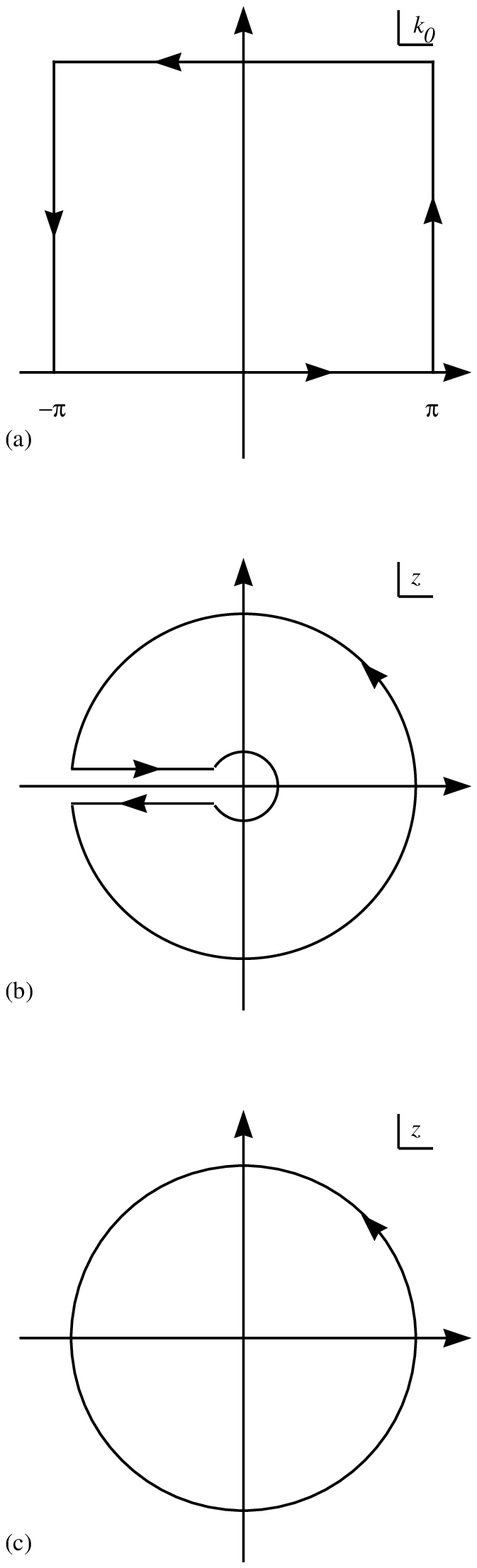}
	\end{center}
	\caption{Contours for integrating over~$k_0$.
	(a)~The complex~$k_0$ plane; 
	(b)~conformal map of onto the complex~$z$ plane; 
	(c)~the final contour in the complex~$z$ plane.}
	\label{contour}
\end{figure}

Consequently, it is prudent, if cumbersome, to integrate over~$k_0$ 
analytically~\cite{Eic90,Dav92,Mor93}.
The easiest way is contour integration.
In a lattice theory, one can proceed as follows, see 
Fig.~\ref{contour}a.
One is to integrate on the real axis over the interval $[-\pi,\pi]$.
At $\pm\pi$ one extends the contour vertically to $\pm\pi+i\infty$.
Then one closes the contour with a segment from $\pi+i\infty$ to 
$-\pi+i\infty$, resulting in a rectangular contour.
 
Since the integrand is a periodic function of $k_0$, the contributions 
from the two vertical sides cancel.
For the Wilson action, the contribution from the top vanishes;
for improved actions, however, this need not be the case; then the 
top must be subtracted out explicitly.
This can be done elegantly by the conformal mapping $z=e^{ik_0}$.
The rectangular contour maps into the contour shown in 
Fig.~\ref{contour}b.
The top of the rectangle maps into the small circle around the origin; 
whether it needs to be subtracted back out, or not, it is always 
correct to take only the outer circle at $|z|=1$.
The final contour is shown in Fig.~\ref{contour}c.

As a function of~$z$ the integrand has poles at physical locations and
at the origin.
The latter restore the contribution from a nonvanishing ``top of the 
rectangle.''
It is then straightforward to integrate over~$k_0$ for all forms of 
$k_0$~dependence appearing in Eqs.~(\ref{F0})--(\ref{FcE2}).

\widetext
\section{Pauli-Villars Functions}
\label{ABPV}
Here we give explicit expressions for the self-energy functions with 
the Pauli-Villars regulator.
They are needed to reconstruct the lattice results from the 
coefficients in the tables.

After introducing Feynman parameters and performing textbook 
manipulations one obtains
\begin{equation}
 	\Sigma^{[1]}_{\text{PV}}(p^2;m)=
 	-\frac{C_F}{16\pi^2}\int_0^1 dx\,(4m+2ix \kern+0.1em /\kern-0.55em p)\;
 		\ln\frac{\Delta_1(x)}{\Delta_\lambda(x)},
	\label{PV Sigma Feynman parameters}
\end{equation}
where
\begin{equation}
 	\Delta_{\lambda}(x)=x\lambda^2 + (1-x)m^2 + x(1-x)p^2.
	\label{PV Delta}
\end{equation}
Carrying out the integration over~$x$ and setting $\lambda\to 0$, one 
finds
\begin{equation}
	A^{[1]}_{\text{PV}}(-m^2;m)=-\frac{C_F}{16\pi^2}\frac{1}{m^2}
	\left[(1-2m^2)\varphi(m^2) + 1 - \ln m^2 \right],
	\label{A PV}
\end{equation}
and, recalling Eq.~(\ref{C=mB}),
\begin{equation}
	B^{[1]}_{\text{PV}}(-m^2;m)=4\frac{C_F}{16\pi^2}\varphi(m^2)
	\label{B PV}
\end{equation}
where
\begin{equation}
\varphi(m^2)=\left\{
\begin{array}{ll}
	m^{-2} \left(\case{1}{2}\ln m^2 +
	\sqrt{1-4m^2}\tanh^{-1}\sqrt{1-4m^2}\right), & m^2<\case{1}{4} 
	\\[1.0em]
	m^{-2} \left(\case{1}{2}\ln m^2 -
	\sqrt{4m^2-1} \tan^{-1}\sqrt{4m^2-1}\right), & m^2>\case{1}{4}
\end{array}\right. .
\end{equation}
As $m^2\to 0$, $\varphi(m^2)\to-1+\ln m^2+ m^2(\case{1}{2}+\ln m^2)$;
as $m^2\to\infty$, $\varphi(m^2)\to-\pi/m$.
Differentiating with respect to~$p_0$, one finds
\begin{equation}
	m\dot{A}^{[1]}_{\text{PV}}(-m^2;m)=-4\frac{C_F}{16\pi^2} 
		\left[\xi(m^2)  + \case{1}{2}\ln(\lambda^2/m^2)\right]
	\label{A dot PV}
\end{equation}
\begin{equation}
	m\dot{B}^{[1]}_{\text{PV}}(-m^2;m)=-8\frac{C_F}{16\pi^2}
		\left[\eta(m^2) + \case{1}{2}\ln(\lambda^2/m^2)\right]
	\label{B dot PV}
\end{equation}
where
\begin{equation}
\xi(m^2)=m^{-2}(1-\ln m^2)-(1-m^{-2})\varphi(m^2) + \psi(m^2),
\end{equation}
\begin{equation}
\eta(m^2)=\ln m^2 - \varphi(m^2) - \psi(m^2),
\end{equation}
with
\begin{equation}
\psi(m^2)=\case{1}{2}\ln m^2 + \left\{
\begin{array}{ll}
	(1-4m^2)^{-1/2}\tanh^{-1}\sqrt{1-4m^2}, & m^2<\case{1}{4} 
	\\[1.0em]
	(4m^2-1)^{-1/2} \tan^{-1}\sqrt{4m^2-1}, & m^2>\case{1}{4}
\end{array}\right. .
\end{equation}
As $m^2\to 0$, $\psi(m^2)\to-m^2(\ln m^2 +1)$;
as $m^2\to\infty$, $\psi(m^2)\to\case{1}{2}\ln m^2$.

\narrowtext


\begin{thebibliography}{9}
\bibitem{Gou97}
B. J. Gough, {\em et al.}, Phys. Rev. Lett. {\bf 79}, 1622 (1997).
\bibitem{Eic87}
E. Eichten, Nucl. Phys. B Proc. Suppl. {\bf 4}, 170 (1988).
\bibitem{Eic90}
E. Eichten and B. Hill, Phys. Lett. {\bf B240}, 193 (1990).
\bibitem{Cas86}
W. E. Caswell and G. P. Lepage, Phys. Lett. {\bf 167B}, 437 (1986).
\bibitem{Lep87}
G. P. Lepage and B. A. Thacker, 
Nucl. Phys. B Proc. Suppl. {\bf 4}, 199 (1988);\\
B. A. Thacker and G. P. Lepage, Phys. Rev. {\bf D43}, 196 (1991);\\
G. P. Lepage, {\em et al.}, Phys. Rev. {\bf D46}, 196 (1992).
\bibitem{KKM97}
A. X. El-Khadra, A. S. Kronfeld, and P. B. Mackenzie,
Phys. Rev. {\bf D55}, 3933 (1997).
\bibitem{Wil77}
K. G. Wilson, in {\em New Phenomena in Subnuclear Physics}, edited by
A. Zichichi (Plenum, New York, 1977).
\bibitem{She85}
B. Sheikholeslami and R. Wohlert, Nucl. Phys. {\bf B259}, 572 (1985).
\bibitem{GHS84}
R. Groot, J. Hoek, and J. Smit, Nucl. Phys. {\bf B237}, 111 (1984).
\bibitem{Gab91}
E. Gabrielli, {\em et al.}, Nucl. Phys. {\bf B362}, 475 (1991);\\
A. Borrelli, C. Pittori, R. Frezzotti, and E. Gabrielli,
Nucl. Phys. {\bf B409}, 382 (1993).
\bibitem{Cap97}
S. Capitani, {\em et al.}, DESY~97-181 ({\tt hep-lat/9709049});
DESY~97-216 ({\tt hep-lat/ 9711007}).
\bibitem{Dav92}
C. T. H. Davies and B. A. Thacker, Phys. Rev. {\bf D45}, 915 (1992).
\bibitem{Mor93}
C. Morningstar, Phys. Rev. {\bf D48}, 2265 (1993);
{\bf D50}, 5902 (1994).
\bibitem{Sin97}
S. Sint and P. Weisz, Nucl. Phys. {\bf B502}, 251 (1997).
\bibitem{Mer94}
A. S. Kronfeld and B. P. Mertens, 
Nucl. Phys. B Proc. Suppl. {\bf 34}, 495 (1994).
\bibitem{Mer95}
A. X. El-Khadra and B. P. Mertens, 
Nucl. Phys. B Proc. Suppl. {\bf 42}, 406 (1995).
\bibitem{Mer97}
B. P. G. Mertens, University of Chicago Ph.~D. Thesis, 
unpublished (1997).
\bibitem{Lep93}
G. P. Lepage and P. B. Mackenzie, Phys. Rev. {\bf D48}, 2265 (1993).
\bibitem{Kro97}
For a preliminary look at $\bar{m}_{\text{ch}}(m_{\text{ch}})$, see
A. S. Kronfeld, FERMILAB-CONF-97/326-T ({\tt hep-lat/9710007}).
\bibitem{Wil74} 
K. G. Wilson, Phys. Rev. {\bf D10}, 2445 (1974).
\bibitem{Kur97}
Y. Kuramashi, KEK-CP-55 ({\tt hep-lat/9705036}).
\bibitem{Aok98}
S. Aoki, S. Hashimoto, K.-I. Ishikawa, and T. Onogi, in preparation.
\end{thebibliography}
\end{document}